\documentclass[preprint,showpacs,preprintnumbers,amsmath,amssymb]{revtex4}
\usepackage{graphicx}
\usepackage{dcolumn}
\usepackage{bm}
\begin{document}
\title{Magnetic properties of RFe$_2$Zn$_{20}$ and RCo$_2$Zn$_{20}$(R = Y, Nd, Sm, Gd - Lu)}
\author{Shuang Jia, Ni Ni, S. L. Bud'ko, P. C. Canfield}
\affiliation{Ames Laboratory, and Department of Physics and Astronomy\\ Iowa State University\\Ames, Iowa 50011, USA\\}

\begin{abstract}
Magnetization, resistivity and specific heat measurements were performed on solution-grown, single crystals of RFe$_2$Zn$_{20}$ and RCo$_2$Zn$_{20}$(R = Y, Nd, Sm, Gd - Lu).
Whereas LuCo$_2$Zn$_{20}$ and YCo$_2$Zn$_{20}$ manifest unremarkable, metallic behavior, LuFe$_2$Zn$_{20}$ and YFe$_2$Zn$_{20}$ manifest behaviors characteristic of nearly ferromagnetic Fermi liquids.
When the well defined $4f$ local moments (Gd$^{3+}$ - Tm$^{3+}$) were embedded into this strongly polarizable host, they all manifest enhanced, ferromagnetic ordering and the values of $T_{\mathrm{C}}$ for RFe$_2$Zn$_{20}$ (R = Gd - Tm) scale with the de Gennes factor.
In addition, the data on the RFe$_2$Zn$_{20}$ compounds indicate a small crystal electronic field (CEF) effect, compare with the interaction energy scale.
On the other hand, the local moment bearing members of RCo$_2$Zn$_{20}$ (R = Nd, Sm, Gd - Tm) manifest weak magnetic interactions and the magnetic properties for R = Dy - Tm members are strongly influenced by the CEF effect on the R ions.
The magnetic anisotropy and specific heat data for the Co series were used to determine the CEF coefficient of R ion in the cubic coordination.
These CEF coefficients, determined for the Co series, are consistent with the magnetic anisotropy and specific heat data for the Fe series, which indicates similar CEF effects for the Fe and Co series.
Such analysis, combined with specific heat and resistivity data, indicates that for R = Tb - Ho, the CEF splitting scale is smaller than their $T_{\mathrm{C}}$ values, whereas for ErFe$_2$Zn$_{20}$ and TmFe$_2$Zn$_{20}$ the $4f$ electron loses part of its full Hund's rule ground state degeneracy above $T_{\mathrm{C}}$.
YbFe$_2$Zn$_{20}$ and YbCo$_2$Zn$_{20}$ manifest typical, but distinct heavy Fermion behaviors associated with different Kondo temperatures.

\end{abstract}


\maketitle
\section{Introduction}

Intermetallic compounds consisting of rare earth and transition metals, as well as metalloids, have versatile magnetic properties.\cite{franse_magnetic_1993, szytula_handbook_1994}
Compounds with itinerant $d$ electrons are of particular interest when they are in the vicinity the Stoner transition: such systems, characterized as nearly or weakly ferromagnet, manifest strongly correlated electronic properties.\cite{moriya_spin_1985}
On the other hand, heavy rare earth ions manifest magnetic versatility associated with the $4f$ electrons: null magnetism (Y$^{3+}$ or Lu$^{3+}$), pure spin, local moment magnetism (Gd$^{3+}$), potentially anisotropic, crystal electric field (CEF) split, local moment magnetism (Tb$^{3+}$ - Tm$^{3+}$), and more exotic magnetism: Yb ions may hybridize with conduction electrons and manifest so-called heavy fermion behavior.
Needless to say, series of materials that combine these interesting versatilities have attracted the attention of physicists.
For example, the binary RCo$_2$ (R = rare earth) compounds, with the nearly ferromagnetic (FM) end members YCo$_2$ and LuCo$_2$, and the local moment, FM members (R = Pr, Nd, Gd - Tm), have been studied for more than 35 years.\cite{duc_itinerant_1999, duc_formation_1999}

\begin{figure}
  \begin{center}
  \includegraphics[clip, width=0.45\textwidth]{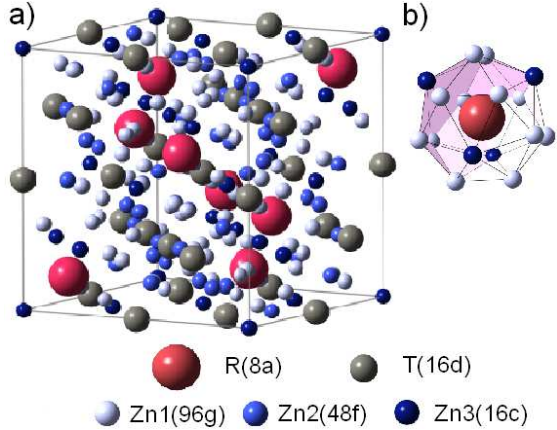}\\
  \caption{(Color online) (a) The cubic unit cell of RT$_2$Zn$_{20}$. (b) The CN16 Frank-Kasper polyhedron of rare earth ions.}
  \label{Fig1poly}
  \end{center}
\end{figure}

Discovered by Nasch \it et. al \rm \cite{nasch_ternary_1997}, the RT$_2$Zn$_{20}$ (R = rare earth, T = transition metal in the Fe, Co and Ni groups) series of compounds crystallize in the cubic CeCr$_2$Al$_{20}$ structure \cite{kripyakevich_RCr2Al20_1968, thiede_euta2al20_1998, moze_crystal_1998}.
The R and T ions occupy their own single, unique, crystallographic sites with cubic and trigonal point symmetry respectively, whereas the Zn ions have three unique crystallographic sites (Fig. \ref{Fig1poly} a).
Both the R and the T ions are fully surrounded by the shells of Zn that consist of the nearest neighbors (NNs) and the next nearest neighbors (NNNs).
This means that there are no R-R, T-T or R-T nearest neighbor and the shortest R-R spacing is $\sim$~6 {\AA}.
For the R ions, a Frank-Kasper polyhedron with coordinate number (CN) 16 is formed from 4 Zn NNs on the $16c$ site and 12 Zn NNNs on the $96g$ site (Fig. \ref{Fig1poly} b).

Recently, studies of the thermodynamic and transport properties of these intermetallics have revealed varied magnetic behavior.\cite{jia_nearly_2007, torikachvili_six_2007, jia_GdY_2007, allGd}
YFe$_2$Zn$_{20}$ and LuFe$_2$Zn$_{20}$ are archetypical examples of nearly ferromagnetic Fermi liquids (NFFL) with Stoner enhancement factors of $Z = 0.88$ (where $\chi _{T=0} = \chi _{Pauli}/(1-Z)$).
By embedding large, Heisenberg type moments associated with Gd$^{3+}$ ions in this highly polarizable medium, GdFe$_2$Zn$_{20}$ manifests highly enhanced FM order.
On the other hand, GdCo$_2$Zn$_{20}$ manifests ordinary, low temperature, antiferromagnetic (AFM) order, correspondent to the `normal metal' behavior of the conduction electron host, YCo$_2$Zn$_{20}$.\cite{jia_nearly_2007}
In addition to these interesting, $3d$ electron and local moment properties, six related YbT$_2$Zn$_{20}$ compounds (T = Fe, Co, Ru, Rh, Os and Ir) show heavy fermion ground states, associated with different Kondo temperatures ($T_{\mathrm{K}}$) and Yb ion degeneracies.\cite{torikachvili_six_2007}

Given the similarities and differences between the RFe$_2$Zn$_{20}$ and RCo$_2$Zn$_{20}$ (R = Gd, Y, Lu) series, it becomes important to study all of the R = Y, Gd - Lu members in detail.
A comparative study of the RFe$_2$Zn$_{20}$ and RCo$_2$Zn$_{20}$ series will help to further understand the magnetic interaction between the local moments by means of the strongly polarizable medium, particularly with the crystal electronic field (CEF) effect associated with non-zero orbital angular momentum. 
Furthermore, given the very similar CN-16 Frank-Kasper polyhedron for R ions, as well as the less than 2{\%} difference of lattice constants for the whole RT$_2$Zn$_{20}$ families, the study of the CEF effect on these local moment members will also help in the understanding of the varied heavy fermion states of YbT$_2$Zn$_{20}$, which were thought to be due to the competition between temperature scales associated with the CEF splitting and the Kondo effect.\cite{torikachvili_six_2007}
 
In this paper, we present the results of magnetization, heat capacity and resistivity measurements on RFe$_2$Zn$_{20}$ and RCo$_2$Zn$_{20}$ (R = Y, Nd, Sm, Gd - Lu) compounds.
Compared with the weakly correlated behaviors for YCo$_2$Zn$_{20}$ and LuCo$_2$Zn$_{20}$, YFe$_2$Zn$_{20}$ and LuFe$_2$Zn$_{20}$, manifest clear, NFFL behaviors associated with the spin fluctuation of the itinerant electrons.
For the RFe$_2$Zn$_{20}$ compounds (R = Gd - Tm), the well-defined, local moment members all manifest enhanced FM ordering with $T_{\mathrm{C}}$ values that roughly scale with the de Gennes factor.
Their anomalous, temperature dependent susceptibility and resistivity can be explained as the result of local moments embedded in a NFFL host.
In contrast, for the RCo$_2$Zn$_{20}$ series, only Gd and Tb members manifest AFM ordering above 2~K, and the magnetic properties for R = Dy - Tm clearly manifest features associated with single ion CEF effects on the R ions in the cubic symmetry coordination.
For the R = Tb - Tm members in the Co series, the CEF parameters can be determined from the magnetic anisotropy and the specific heat data, and are roughly consistent with the results of calculations using a point charge model.
For the Fe series, the R = Tb - Tm members show moderate magnetic anisotropy in their ordered states, mainly due to the CEF effect on the R ions, which is consistent with the magnetic anisotropy found for the Co members.
These results, as well as the analysis of the heat capacity and resistivity data, indicate that the FM state emerges from the fully degenerate Hund's rule ground state for RFe$_2$Zn$_{20}$ (R = Gd - Ho), whereas ErFe$_2$Zn$_{20}$ and TmFe$_2$Zn$_{20}$ manifests CEF splitting above their Curie temperatures.

\section{Experimental Methods}

\begin{figure}
  \begin{center}
  \includegraphics[clip, width=0.45\textwidth]{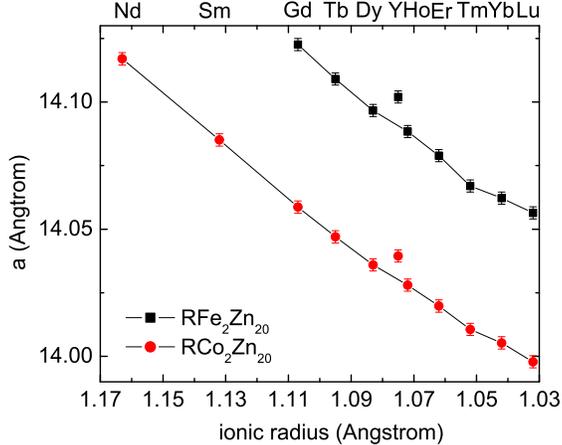}\\
  \caption{(Color online) The lattice constants ($a$) for RFe$_2$Zn$_{20}$ and RCo$_2$Zn$_{20}$ versus the radius of the trivalent rare earth ion with CN = 9. \cite{shannon_ion} The error bars were estimated from the standard variation of four seperate measurement made on one batch of sample.}
  \label{Fig2lattice}
  \end{center}
\end{figure}

Single crystals of RFe$_2$Zn$_{20}$ and RCo$_2$Zn$_{20}$ (R = Y, Gd - Lu) were grown from high temperature, Zn rich solutions with the initial concentration of starting elements being R:(Fe/Co):Zn = 2:4:94, as described previously \cite{canfield_growth_1992, jia_nearly_2007}.
The small amount of residual Zn on the resulting crystals was removed by submerging the crystals into an ultrasonic bath of 0.5 vol. \% HCl in H$_2$O for 0.5 - 1 hour.
Attempts were made to grow even lighter rare earth members (R = Eu, Nd and Ce) for the Fe series, but no ternary compound was found to form.
On the other hand, for the Co series, single crystals for the R = Nd and Sm members were grown using the same growth method. 
Room temperature, powder X-ray diffraction measurements, using Cu $K_{\alpha}$ radiation, were performed on the samples with additional Si powder ($a=5.43088$~{\AA}) used as standard.
Figure \ref{Fig2lattice} shows the lattice constants for the two series of compounds, obtained by using the Rietica, Rietveld refinement program, plotted with respect to the effective radius of R$^{3+}$ with CN = 9 \cite{shannon_ion}, since the data is absent for larger CN.
The variation of the lattice constant manifests the well-known, lanthanide contraction for R = Gd - Lu with no evident deviation for R = Yb.
However, the relatively larger lattice constants for YFe$_2$Zn$_{20}$ and YCo$_2$Zn$_{20}$ indicate that, with this large CN, the effective ionic radii of Y$^{3+}$ is between Tb$^{3+}$ and Dy$^{3+}$ rather than between Dy$^{3+}$ and Ho$^{3+}$.
This deviation for Y$^{3+}$ ions is not unprecedented in the isostructural compounds RRu$_2$Zn$_{20}$ \cite{nasch_ternary_1997} and RMn$_2$In$_x$Zn$_{20-x}$ \cite{benbow_rmn2zn20}, as well as the similar structure compounds RCo$_2$ \cite{pearsons}.
Additional single crystal X-ray diffraction measurements were preformed on R = Gd, Tb, Er and Lu members of the $\mathrm{RFe_2Zn_{20}}$ series and demonstrated full occupancy on all crystallographic sites (within the detection errors) and the same lattice as the powder X-ray values.\cite{Xray}
This result is complicated by the difficulty in resolving the difference between Fe and Zn elements with very similar X-ray scattering strengths.

It is worth noting though, that for the local moment bearing members in Fe series, single crystals obtained from different ratios of starting element concentrations manifest detectably different magnetic ordering temperatures.
These differences, tentatively associated with very subtle variations of element occupancy on the crystallographic sites\cite{neutron}, are related to the extreme sensitivity to the small disorder for compounds with such strongly correlated conduction electron backgrounds.
A detail discussion of this is presented in Appendix A.

Measurements of electrical resistivity  were preformed by using a standard AC, four-probe technique.
The samples were cut into bars, with typical lengths of 2--3 mm, along their crystallographic [110] direction.
These bars were measured in a Quantum Design physical properties measurement system, PPMS-14 and/or PPMS-9 instruments ($T =1.85$ -- $310$ K) with $f = 16$ Hz, $I = 3$ -- $0.3$ mA.
Temperature dependent specific heat measurements were also performed in these PPMS units, using the heat capacity option.
In order to study the magnetic part of specific heat, the contribution from the conduction electrons and lattice vibrations was subtracted.
This was performed by subtracting off the specific heat of the non-magnetic members YT$_2$Zn$_{20}$ or LuT$_2$Zn$_{20}$ [$C_{mag}=C_p(\mathrm{RT_2Zn_{20}})-C_p(\mathrm{Lu/YT_2Zn_{20}})$].
The magnetic part of entropy can then be estimated as $S_M=\int \frac{C_{p}(\mathrm{RT_2Zn_{20}})-C_{p}(\mathrm{Lu/YT_2Zn_{20}})}{T}\, \mathrm{d}T$, in which the specific heat data below 1.8~K (0.4~K for $\mathrm{TmCo_2Zn_{20}}$) were approximated by a linear extrapolation to $C_p = 0$ at 0~K.

Magnetization measurements were carried out in Quantum Design magnetic properties measurement system (MPMS), superconducting quantum interface device (SQUID) magnetometers in varied applied fields ( $H \leq  55$ kOe) and temperatures ($T = 1.85$~--~$375$ K).
In the measurements of magnetization for the FM-ordered, RFe$_2$Zn$_{20}$ compounds, the effects of the demagnetizing field are relatively small, because of the dilute nature of the magnetic moments:\cite {allGd, jia_GdY_2007} the maximum demagnetizing field is about 3000 Oe for a plate-like shaped sample.
Even so, in order to obtain accurate magnetization isotherms near $T_{\mathrm{C}}$, the samples were cut and measured with the applied field along their long axis so as to minimize this already small demagnetization effect. 

\section{Results}

We start characterizing the compounds with the non-magnetic rare earth ions of the series: Y(Lu)Fe$_2$Zn$_{20}$ and Y(Lu)Co$_2$Zn$_{20}$.
Without any $4f$ electronic magnetism, these compounds manifest the electronic and magnetic properties associated with the conduction electron background of each series.
Next, we will introduce the two series of compounds with well-defined $4f$ local moments: R = Gd - Tm.
We will examine the magnetization and specific heat data for the Co series at first.
Then an overview of the magnetic properties for the Fe series will be presented next.
After that, the magnetization, specific heat and resistivity data will be presented for each Fe member separately.
Finally, similar data for the R = Yb heavy fermion compounds, YbFe$_2$Zn$_{20}$ and YbCo$_2$Zn$_{20}$ will be presented.
 
\subsection{Y(Lu)Fe$_2$Zn$_{20}$ and Y(Lu)Co$_2$Zn$_{20}$}

\begin{figure}
  \begin{center}
  \includegraphics[clip, width=0.45\textwidth]{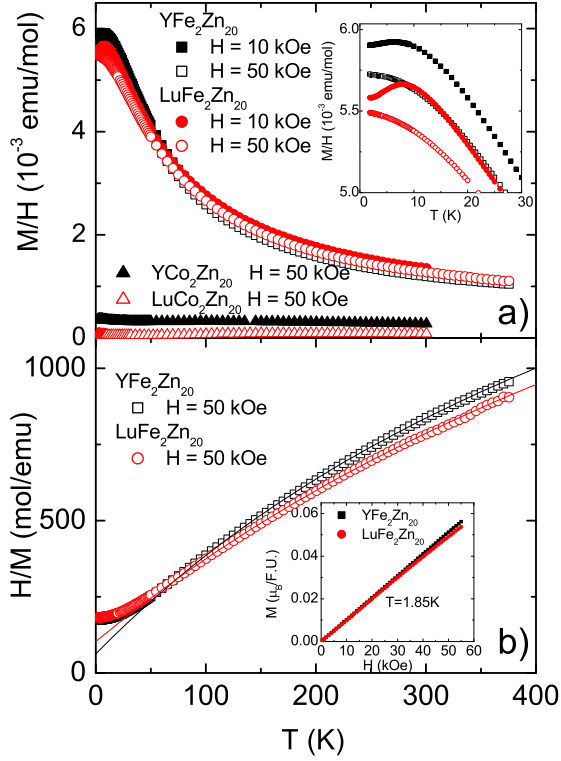}\\
  \caption{(Color online) (a) Temperature dependent magnetization $M$ divided by the applied field $H$ for YFe$_2$Zn$_{20}$ and LuFe$_2$Zn$_{20}$ as well as their Co analogues for $H = 10$ kOe and 50 kOe. Inset: a blow-up plot at low temperature. (b) $H/M$ for YFe$_2$Zn$_{20}$ and LuFe$_2$Zn$_{20}$. The solid lines present the modified Curie-Weiss [$\chi (T)=C/(T-\theta _C)+\chi _0$] fit for the data above 100~K. Inset: field dependent magnetization at 1.85~K.}
  \label{FigYLu1M}
  \end{center}
\end{figure}

Temperature dependent magnetization data (divided by the applied field) for Y(Lu)Fe$_2$Zn$_{20}$ and Y(Lu)Co$_2$Zn$_{20}$ are shown in Fig. \ref{FigYLu1M} a.
The Fe members manifest similar, strongly enhanced, temperature-dependent paramagnetic signals, whereas the Co members manifest essentially temperature-independent, Pauli paramagnetic signals.
The low temperature features for the two Fe compounds are shown in the inset of Fig. \ref{FigYLu1M}.
In the applied field of 10~kOe, the magnetization signals of both Fe members show a faint maximum below 10 K, whereas the high magnetic field (50 kOe) suppresses the lowest temperature $M/H$ values, as well as the maximum.
In our experience on the measurements of different batches of samples, these low temperature features are moderately sample-dependent (different samples may show 20{\%} different magnetization signal and 2--3 K difference in the temperature of the maximum, $T_{max}$).
Nevertheless, the maximum of the temperature dependent susceptibility, $\chi (T)$, is a common feature in the NFFLs, such as Pd \cite{pd_kia}, YCo$_2$ and LuCo$_2$ \cite{yco2_kia}, as well as TiBe$_2$ \cite{tibe2_kia}, although quantitative calculation of $\chi (T)$ still presents a challenge even for the simplest case of Pd \cite{zellermann_onsagerpd, larson_pd}.
The field suppression of the magnetization (and $T_{max}$) at low temperature is not attributed to the possible existence of a paramagnetic impurity contribution (which would contribute more to the value of $M/H$ at lower temperature and lower field, and therefore suppress the maximum of $M/H$ in lower field), but, as discussed below, to the intrinsic variation of $\chi =dM/dH$  with respect to $H$ at different temperatures.

Figure \ref{FigYLu1M} b shows that above a characteristic temperature ($T^{\ast }\sim 50$~K), the susceptibility of YFe$_2$Zn$_{20}$ and LuFe$_2$Zn$_{20}$ can be approximately fitted by a Curie-Weiss (CW) term [$\chi (T)=C/(T-\theta _C)$] plus a temperature-independent term ($\chi _0 $).
The values of effective moment ($\mu _{eff}$), $\theta _C$ and $\chi _0 $ are extracted as 1.0 $\mu _B$/Fe, -16 $K$, $3.8\times 10^{-4}$emu/mol and 1.1 $\mu _B$/Fe, -33 $K$, $3.4\times 10^{-4}$emu/mol for YFe$_2$Zn$_{20}$ and LuFe$_2$Zn$_{20}$, respectively.
These values of $\mu _{eff}$ are significantly larger than the estimated induced moment of Fe site in the FM ground state of GdFe$_2$Zn$_{20}$, $\sim 0.35 \mu _B$/Fe. \cite{jia_nearly_2007}
Such apparent CW-like behavior was also observed in other NFFL systems. \cite{shimizu_pdhight,yco2_kia}
In the context of the spin fluctuation model \cite{moriya_spin_1985}, itinerant electronic systems can manifest CW-like behavior with a Curie constant related to the local amplitude of the spin fluctuation.
The magnetization data at the base temperature (1.85 K) show nearly linear dependent with the applied field (Inset in Fig. \ref{FigYLu1M} b), which is distinct from the Brillouin function type of magnetization curves associated with local moments. 

\begin{figure}
  \begin{center}
  \includegraphics[clip, width=0.45\textwidth]{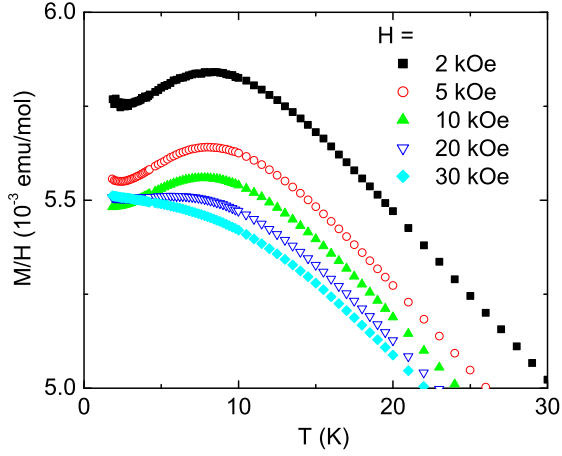}\\
  \caption{$M/H$ for LuFe$_2$Zn$_{20}$. From right to left: $H = 2$~kOe, 5~kOe, 10~kOe, 20~kOe and 30~kOe.}
  \label{FigLu1}
  \end{center}
\end{figure}

\begin{figure}
  \begin{center}
  \includegraphics[clip, width=0.45\textwidth]{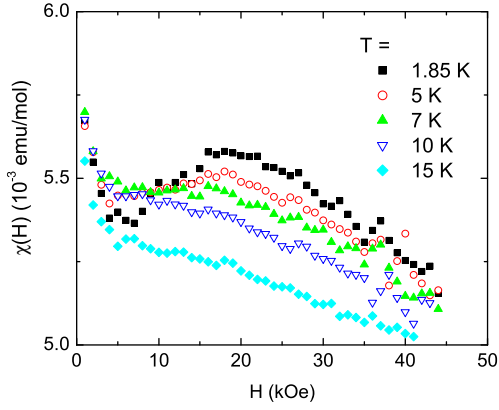}\\
  \caption{$\chi (H)$ for LuFe$_2$Zn$_{20}$ at varied temperature.}
  \label{FigLu2}
  \end{center}
\end{figure}

In order to better understand the variation of the maximum in temperature dependent $M/H$ data for YFe$_2$Zn$_{20}$ and LuFe$_2$Zn$_{20}$, $M(T)$ and $M(H)$ measurements were performed on LuFe$_2$Zn$_{20}$ for varied applied field and temperature respectively.
Figure \ref{FigLu1} shows that the magnetic field suppresses the values of $M/H$, as well as the maximum of $M/H$, which disappears when $H \geq 20$~kOe.
The values of the field dependent susceptibility, $\chi (H)$, were extracted as  $\chi (H) = \frac{M(H+\Delta H)-M(H)}{\Delta H}$ with $\Delta H = 1000$~Oe from the $M(H)$ data at varied temperature (Figure \ref{FigLu2}).
For $T \geq  10$~K, the values of $\chi (H)$ monotonically decrease with increase $H$, whereas a local maximum appears around $20$~kOe in the data sets as $T \leq 7$~K. 
This critical temperature ($\sim 7$~K) is close to the $T_{max}$; the maximum of $\chi (H)$ ($H = 20$~kOe) is also close to the suppression field determined by Fig. \ref{FigLu1}.
This curious, field dependent, susceptibility at varied temperature is reminiscent to the one of TiBe$_2$, albeit the amplitude of local maximum in $\chi (H)$ is much smaller. \cite{tibe2_kia, acker_enhanced_1981}
In the case of TiBe$_2$, the reason of anomalous field-dependent magnetization is also still not clear. \cite{jeong_fermi_2006}

\begin{figure}
  \begin{center}
  \includegraphics[clip, width=0.45\textwidth]{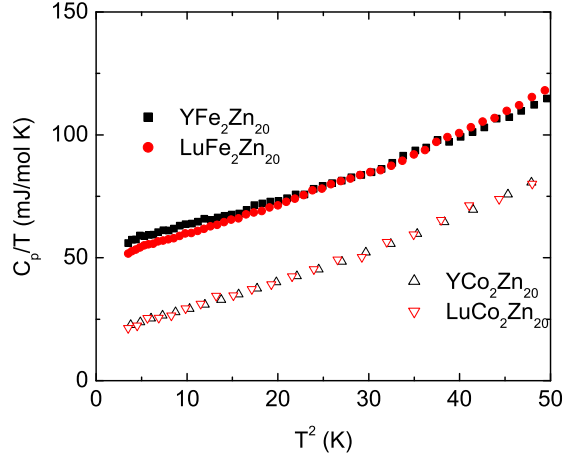}\\
  \caption{(Color online) Low temperature specific heat data of YFe$_2$Zn$_{20}$ and LuFe$_2$Zn$_{20}$ (plotted as $C_p/T$ versus $T^2$), as well as the Co analogues.}
  \label{FigYLu2Cp}
  \end{center}
\end{figure}

Figure \ref{FigYLu2Cp} presents the low temperature specific heat data for YFe$_2$Zn$_{20}$ and LuFe$_2$Zn$_{20}$, as well, as for the Co analogues, plotted as $C_p/T$ versus $T^2$.
All four compounds manifest clear Fermi liquid behavior ($C_p=\gamma T+\beta T^3$).
The similar $\beta $ values (represented as the slopes of the data sets in the plot, $\sim 1.2 \mathrm{mJ/mol K}^4$) indicate the similar Debye temperatures for these 4 compounds ($\sim 340$~K \cite{allGd}), consistent with their similar molar mass, similar composition and similar lattice parameters.
On the other hand, the over 2.5 times larger values of electronic specific heat ($\gamma $) of the Fe members indicate a larger density of states at Fermi level [$N(E_f)$], compared to the Co analogues (consistent with the band structure calculation results \cite{allGd}).

The values of the electronic specific heat and magnetic susceptibility can be employed to estimate the Stoner enhancement factor, $Z$, in the context of the Stoner theory: that is, the static susceptibility is enhanced by $\frac{1}{1-Z}$ , whereas the electronic specific heat is not. \cite{jia_nearly_2007}
The estimated $Z$ values of YFe$_2$Zn$_{20}$ and LuFe$_2$Zn$_{20}$ are 0.88, 0.89, respectively, comparable with the estimated values of the canonical NFFL systems: Pd: 0.83, and YCo$_2$: 0.75.\cite{pdyco2Z}

\begin{figure}

  \begin{center}
  \includegraphics[clip, width=0.45\textwidth]{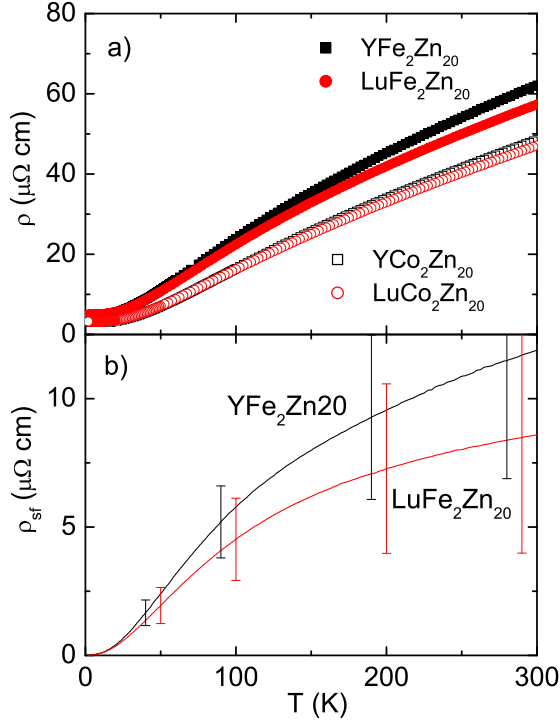}\\
  \caption{(Color online) (a): Temperature dependent resistivity of YFe$_2$Zn$_{20}$ and LuFe$_2$Zn$_{20}$, as well as their Co analogues. (b): estimated spin fluctuation contribution to the resistivity for YFe$_2$Zn$_{20}$ and LuFe$_2$Zn$_{20}$. The error bars were estimated as $\pm 10\%$ of the values of the resistivity for YCo$_2$Zn$_{20}$ and LuCo$_2$Zn$_{20}$ respectively.}
  \label{FigYLu3rou}
  \end{center}
\end{figure}

The temperature dependent electrical resistivity data for YFe$_2$Zn$_{20}$ and LuFe$_2$Zn$_{20}$ are larger than that for the Co analogues over the whole temperature range (Fig. \ref{FigYLu3rou} a).
This is not unexpected for a NFFL since the spin fluctuations will affect the scattering process of the conduction electrons, which leads to an additional contribution to the resistivity.
In order to study the spin fluctuation contribution to the resistivity, the total electrical resistivity $\rho (T)$ is assumed to be:
\begin{equation}
\rho (T)=\rho _{0}+\rho _{ph}(T)+\rho _{sf}(T),
\label{eqn:1}
\end{equation}
where the first, second and third terms represent residual, phonon and spin fluctuation reisitivity, respectively. Assuming the phonon scattering contribution, $\rho _{ph}(T)$, is essentially same for the Fe and Co analogues (as suggested by the similar $\beta $ terms), then, the spin fluctuation scattering contribution, $\rho _{sf}(T)$, can be estimated as: 

\begin{equation}
\rho _{sf}(T)=(\rho -\rho _{0})_{\mathrm{Y/LuFe_2Zn_{20}}}-(\rho -\rho _{0})_{\mathrm{Y/LuCo_2Zn_{20}}}.
\label{eqn:2}
\end{equation}

Shown in Fig. \ref{FigYLu3rou}b, $\rho _{sf}(T)$ for these two compounds increases with temperature and is close to a saturated value (10~$\mu \Omega cm$) at 300~K, within the accuracy of the measurements. 

\begin{figure}
  \begin{center}
  \includegraphics[clip, width=0.45\textwidth]{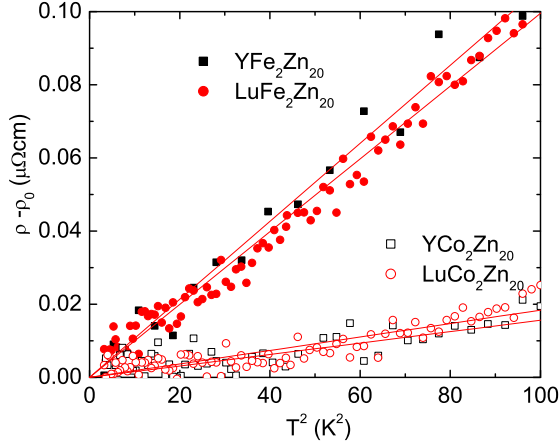}\\
  \caption{(Color online) $\rho $ versus $T^2$ for YFe$_2$Zn$_{20}$ and LuFe$_2$Zn$_{20}$, as well as their Co analogues. The solid lines present the linear fit of the data sets from 2 K to 9 K.}
  \label{FigYLu4rouT2}
  \end{center}
\end{figure}

The analysis of the low temperature resistivity data reveals a quadratic, standard Fermi liquid, behavior [$\rho (T)=\rho _{0}+AT^2$] for all 4 compounds (Fig. \ref{FigYLu4rouT2}).
The $A$ values of the Fe compounds are about 7 times larger than the two Co analogues.
This result is consistent with the 2.5 times larger $\gamma $ values of the Fe compounds, in the context of the Fermi liquid theory, meaning $A$ is proportional to the square of the effective mass of the quasi-particles due to the strong correlation effect, whereas $\gamma $ is proportional to the effective mass.
In the point of view of spin fluctuation theory, nearly FM metals manifest Fermi liquid behaviors at low temperature region with $A$ values enhanced by spin fluctuations.\cite{moriya_spin_1985} 

\subsection{RCo$_2$Zn$_{20}$ (R = Nd, Sm, Gd - Tm)}

\begin{figure}
  \begin{center}
  \includegraphics[clip, width=0.45\textwidth]{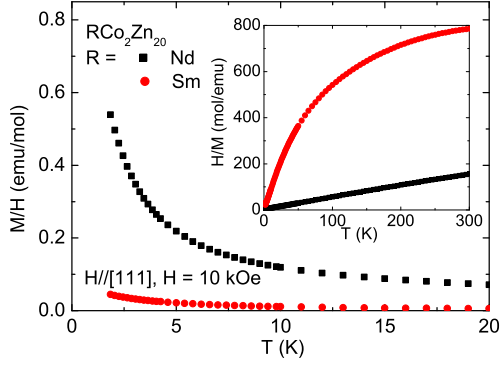}\\
  \caption{(Color online) Temperature dependent magnetization of RCo$_2$Zn$_{20}$ (R = Nd and Sm) compounds, divided by applied field $H = 10000$~Oe. Inset: applied field ($H = 10000$~Oe) divided by the magnetizations of RCo$_2$Zn$_{20}$ (R = Nd and Sm) as a function of temperature.}
  \label{FigNdSm1}
  \end{center}
\end{figure}

Before discussing the heavy rare earth compounds (R = Gd - Yb), the results of thermodynamic measurement on NdCo$_2$Zn$_{20}$ and SmCo$_2$Zn$_{20}$ will be briefly examined.
Figure \ref{FigNdSm1} shows the temperature dependent magnetization data (divided by the applied field $H=1000$ Oe) for NdCo$_2$Zn$_{20}$ and SmCo$_2$Zn$_{20}$.
Neither of them manifest any sign of magnetic ordering above 2 K.
The temperature dependent $H/M$ for NdCo$_2$Zn$_{20}$ shows a CW behavior [$\chi (T)=C/(T-\theta _C)+\chi _0$] with $\mu _{eff}=3.7\mu _B$, $\theta _C=-2.3$ K and $\chi _0=6.8\times 10^{-4}$ emu/mol.
The value of the effective moment is close to the theoretical values for the Hund's rule ground state of the $4f$ electrons of Nd$^{3+}$ ion ($3.6\mu _B$). 
On the other hand, though the magnetization of SmCo$_2$Zn$_{20}$ drops with increase temperature, it does so in a distinctly non-CW manner. 
This behavior is not unexpected in Sm containing compounds \cite{myers_RAgSb2_1999}, and is most likely due to the thermal population of the first excited Hund's rule multiplet.
 
\begin{figure}
  \begin{center}
  \includegraphics[clip, width=0.45\textwidth]{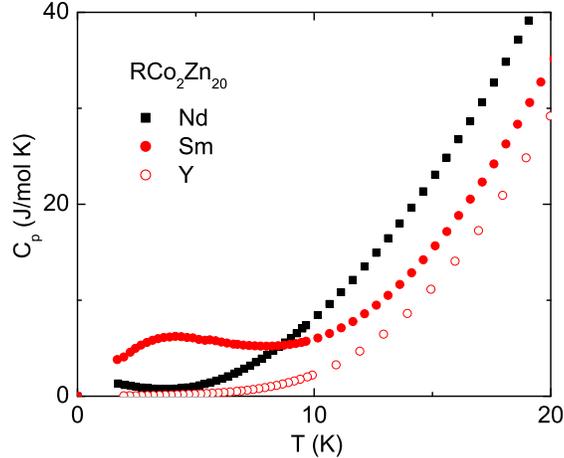}\\
  \caption{(Color online) Temperature dependent specific heat for RCo$_2$Zn$_{20}$ (R = Nd, Sm and Y).}
  \label{FigNdSm2}
  \end{center}
\end{figure}

Specific heat data for NdCo$_2$Zn$_{20}$ and SmCo$_2$Zn$_{20}$ are shown in Fig. \ref{FigNdSm2} along with data for YCo$_2$Zn$_{20}$ for comparison.
The low temperature upturn in the NdCo$_2$Zn$_{20}$ data below 2 K may be due to a lower temperature magnetic ordering or a Schottky anomaly due to the CEF splitting.
The specific heat data for SmCo$_2$Zn$_{20}$ manifest a broad peak around 4 K, which is most likely due to the CEF splitting of the Hund's rule ground state of Sm$^{3+}$.
Both NdCo$_2$Zn$_{20}$ and SmCo$_2$Zn$_{20}$ data increase much faster above 10~K, and remain more than 10 J/mol K larger above 25~K (not shown here), compared with the data for the non-magnetic analogue YCo$_2$Zn$_{20}$. 
On the other hand, the calculated results of the CEF splitting for the Hund's rule ground state of Nd$^{3+}$ ion in a point charge model show the splitting energy levels within 25~K (see Table \ref{table3} below).
This large difference indicates that, at this point, the magnetic part of $C_p$ for NdCo$_2$Zn$_{20}$ and SmCo$_2$Zn$_{20}$ cannot be well estimated, since the $C_p$ data of YCo$_2$Zn$_{20}$ is not an adequate approximation of the non-magnetic background and unfortunately the LaCo$_2$Zn$_{20}$ analogue did not form. 

\begin{figure}
  \begin{center}
  \includegraphics[clip, width=0.45\textwidth]{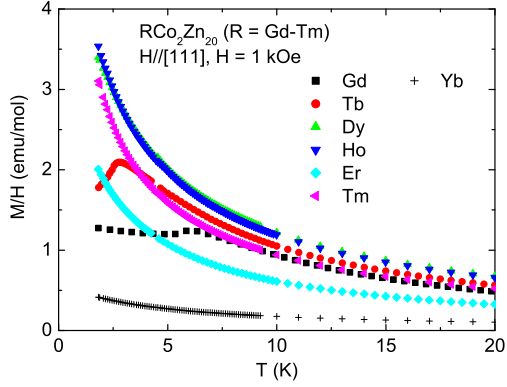}\\
  \caption{(Color online) Temperature dependent magnetization of RCo$_2$Zn$_{20}$ (R = Gd - Yb) compounds, divided by applied field $H = 1000$~Oe.}
  \label{FigAllCo1MH}
  \end{center}
\end{figure}

Temperature dependent magnetization data (divided by the applied field $H = 1000$ Oe) for RCo$_2$Zn$_{20}$ (R = Gd - Yb) are presented in Fig. \ref{FigAllCo1MH}.
In addition to the previously reported, AFM ordered GdCo$_2$Zn$_{20}$ with the Ne\'{e}l temperature $T_{\mathrm{N}}=5.7\pm 0.1$~K\cite{jia_nearly_2007,allGd}, TbCo$_2$Zn$_{20}$ also shows AFM ordering with $T_{\mathrm{N}}=2.5\pm 0.1$~K, which also clearly manifests itself in the specific heat data (shown below in Fig. \ref{FigAllCo3Cp}).
The rest of the members (R = Dy - Yb) do not show magnetic ordering above 2~K.
Due to the relatively low density of state at Fermi level [$N(E_F)$] for the Y and Lu analogues\cite{jia_nearly_2007,allGd} and large R-R separation, such low temperature magnetic ordering for the $4f$ local moments coupled via the Ruderman-Kitter-Kasaya-Yosida (RKKY) interaction is not unexpected.

\begin{figure}

  \begin{center}
  \includegraphics[clip, width=0.45\textwidth]{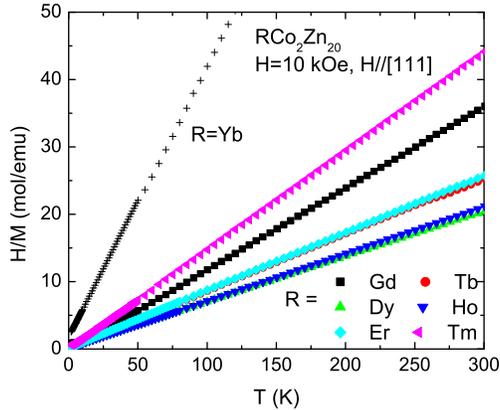}\\
  \caption{(Color online) Applied field ($H = 10000$~Oe) divided by the magnetizations of RCo$_2$Zn$_{20}$ (R = Gd - Yb) as a function of temperature.}
  \label{FigAllCo2HM}
  \end{center}
\end{figure}

Figure \ref{FigAllCo2HM} shows the temperature dependent $H/M$ for R = Gd - Tm and Yb members of the RCo$_2$Zn$_{20}$ series.
All the members, including YbCo$_2$Zn$_{20}$, manifest CW behavior [$\chi (T)=C/(T-\theta _C)+\chi _0$] with negligible small $\chi _0$ ($\leq 2\times 10^{-3}$emu/mol) and the values of $\mu _{eff}$ close to the theoretical values for the Hund's ground state of the $4f$ electronic configurations; all the values of $\theta _C$ are close to 0, consistent with the low magnetic ordering temperatures (Table \ref{table1}). 

\begin{table}
\caption{\label{table1} Paramagnetic Curie temperature, $\theta _C$ (with $\pm 0.1$~K errors) and effective moment, $\mu _{eff}$ [from the CW fit of $\chi (T)$ from 50 K to 300 K]; Ne\'{e}l temperature, $T_{\mathrm{N}}$ for RCo$_2$Zn$_{20}$ compounds (R = Nd, Gd - Yb).}
\begin{ruledtabular}
\begin{tabular}{lcccccccc}
 & Nd & Gd & Tb & Dy & Ho & Er & Tm & Yb \\
\hline
$\theta _C$, K & -2.3 & 3.3 & -2.6 & -3.7 & 1.4 & -2.1 & -0.9 & -5.2\\
$\mu _{eff}$, $\mu _{B}$ & 3.7 & 8.1 & 9.8 & 10.9 & 10.7 & 9.7 & 7.4 & 4.5\\
$T_{\mathrm{N}}$, K & & 5.7 & 2.5 &  &  &  &  & \\
\end{tabular}
\end{ruledtabular}
\end{table}

\begin{figure}

  \begin{center}
  \includegraphics[clip, width=0.45\textwidth]{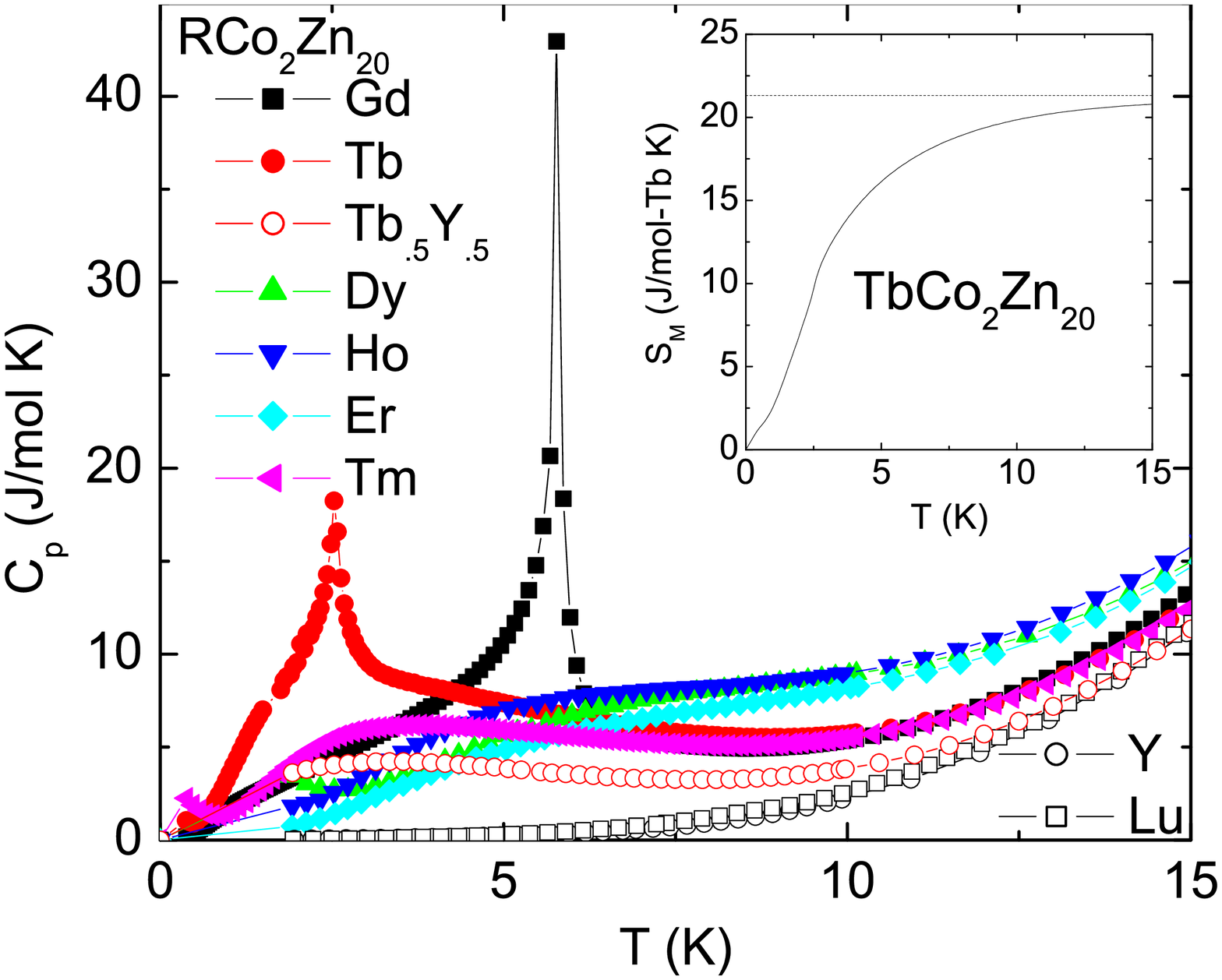}\\
  \caption{(Color online) Temperature dependent specific heat for RCo$_2$Zn$_{20}$ (R = Gd - Tm, Y and Lu), as well as Tb$_{0.5}$Y$_{0.5}$Co$_2$Zn$_{20}$. Inset: temperature dependent magnetic entropy for TbCo$_2$Zn$_{20}$. The dashed line presents the entropy of the full Hund's ground state of Tb$^{+3}$.}
  \label{FigAllCo3Cp}
  \end{center}
\end{figure}

The specific heat data for RCo$_2$Zn$_{20}$ (R = Gd - Tm , Y and Lu), as well as the pseudo-ternary compound Tb$_{0.5}$Y$_{0.5}$Co$_2$Zn$_{20}$ are presented in Fig. \ref{FigAllCo3Cp}.
In addition to the previously studied GdCo$_2$Zn$_{20}$, the specific heat data for TbCo$_2$Zn$_{20}$ manifests a $\lambda $-type of anomaly with a peak position at 2.5~K, the AFM ordering temperature.
In addition to this peak, the $C_p$ data also show a broad shoulder above 2.5~K, which is due to the CEF splitting above the magnetic ordering temperature.
This anomaly, associated with CEF splitting of the $4f$ electronic configuration of Tb$^{3+}$, manifests itself more clearly in the $C_p$ data for Tb$_{0.5}$Y$_{0.5}$Co$_2$Zn$_{20}$: when $T_{\mathrm{N}}$ is suppressed to well below 2 K, the $C_p$ data show a Schottky anomaly with a peak position $\sim 3$~K. 
The magnetic part of entropy for TbCo$_2$Zn$_{20}$ is shown in the inset to Fig. \ref{FigAllCo3Cp}.
Approximately 50 {\%} of the total magnetic entropy is recovered by $T_{\mathrm{N}}$, and by 15~K the full $S_M=R\ln {13}$ is recovered ($R$ is gas constant).
This is consistent with a very small, total CEF splitting in the these compounds, associated with the highly symmetric environment of the R ions.
For the rest of the members, R = Dy - Tm, the specific data show broad, Schottky-type of anomaly below 10~K, as shown in the insets of Fig. \ref{FigDyCo}, \ref{FigHoCo}, \ref{FigErCo} and \ref{FigTmCo} (shown below).
The low temperature upturn for DyCo$_2$Zn$_{20}$ data below 2~K may indicate a magnetic ordering at lower temperature, whereas the upturn for TmCo$_2$Zn$_{20}$ data below 0.7~K may be associated with a magnetic ordering at very low temperature and/or a nuclear Schottky anomaly.

\begin{figure}

  \begin{center}
  \includegraphics[clip, width=0.45\textwidth]{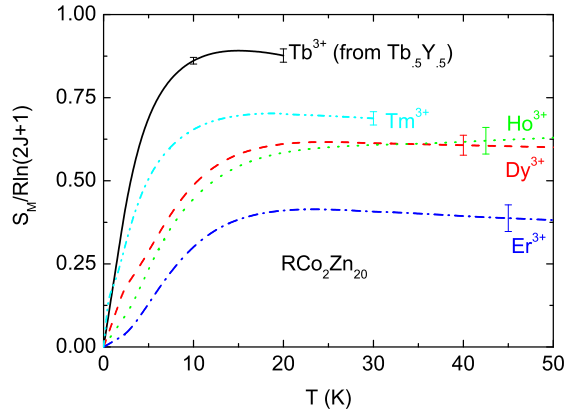}\\
  \caption{(Color online) Normalized magnetic part of entropy for RCo$_2$Zn$_{20}$ (R = Dy - Tm) as well as for Tb$_{0.5}$Y$_{0.5}$Co$_2$Zn$_{20}$ (in units of per mole R$^{3+}$). The error bars were estimated from the $\pm 1$ {\%} of the total entropy.}
  \label{FigallCoSm}
  \end{center}
\end{figure}

The released, magnetic part of entropy above 2~K (above 0.4~K for the TmCo$_2$Zn$_{20}$) are shown in Fig. \ref{FigallCoSm}.
For R = Dy - Tm, there is an obvious deficit of magnetic entropy compared with the value associated with fully degenerated Hund's ground state, which indicates unaccounted entropy below 2~K (0.4~K for TmCo$_2$Zn$_{20}$) associated with low lying CEF levels and magnetic ordering.

In order to better understand the magnetic properties for R = Tb - Tm members, the CEF effect acting on the R ions is evaluated by thermodynamic measurements.
The single-ion Hamiltonian for the R$^{3+}$ is assumed to be the sum of the CEF term, an exchange interaction term and an external field term:
\begin{equation}
\mathcal{H} = \mathcal{H}_{CEF}+\mathcal{H}_{exc}+\mathcal{H}_{ext}.
\label{eqn:3}
\end{equation}
where $\mathcal{H}_{ext}=g_J \mu _B \vec{J}\cdot \vec{H}$, $g_J$ is Lande factor, $\vec{J}$ is the total angular momentum, and $\vec{H}$ is the external magnetic field.

Since the rare earth ions are located in a cubic point symmetry, the CEF term, $H_{CEF}$, can be written as:
\begin{equation}
\mathcal{H}_{CEF}=B^0_4(O^0_4+5O^4_4)+B^0_6(O^0_6-21O^4_6).
\label{eqn:4}
\end{equation}
where $O^m_l$ operators are the well-known Stevens operators \cite{stevens_matrix_1952}, and $B^0_4$ and $B^0_6$ are CEF parameters \cite{lea_cef}. If one follows the work of Lea \it et al. \rm \cite{lea_cef}, this expression can be written as:
\begin{equation}
\mathcal{H}_{CEF}=W[\frac{x}{F4}(O^0_4+5O^4_4)+\frac{1-\left| x \right|}{F6}(O^0_6-21O^4_6)].
\label{eqn:5}
\end{equation}
where $F4$ and $F6$ are factors introduced by Lea \it et al. \rm \cite{lea_cef} and dependent with $J$, $W$ is the energy scale, and $x$ represents the relative importance of the 4th and 6th order terms.

Noticing that the possible magnetic ordering temperatures are below 2 K for RCo$_2$Zn$_{20}$ (R = Dy - Tm), as well as for Tb$_{0.5}$Y$_{0.5}$Co$_2$Zn$_{20}$, the exchange interaction term will be approximated as zero, an approximation that will be better for R = Tm than for R = Dy.
Thus, the CEF parameters for different R ions were determined by fitting the magnetization at 2 K and the temperature dependent specific heat data.

\begin{figure}
  \begin{center}
  \includegraphics[clip, width=0.45\textwidth]{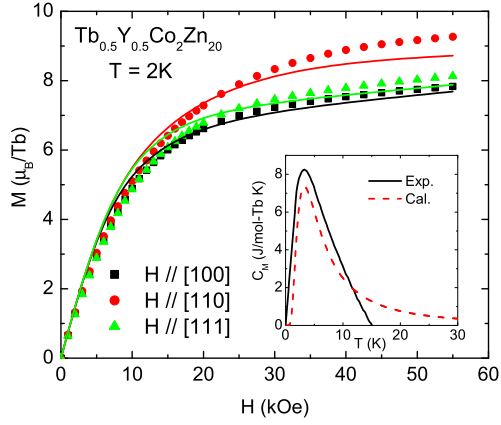}\\
  \caption{(Color online) Field dependent magnetization for Tb$_{0.5}$Y$_{0.5}$Co$_2$Zn$_{20}$ along three principle axes. The solid lines present the fitting results. Inset: magnetic part of specific heat. The solid and dashed line present the experimental and calculated result respectively.}
  \label{FigTbCo}
  \end{center}
\end{figure}

\begin{figure}
  \begin{center}
  \includegraphics[clip, width=0.45\textwidth]{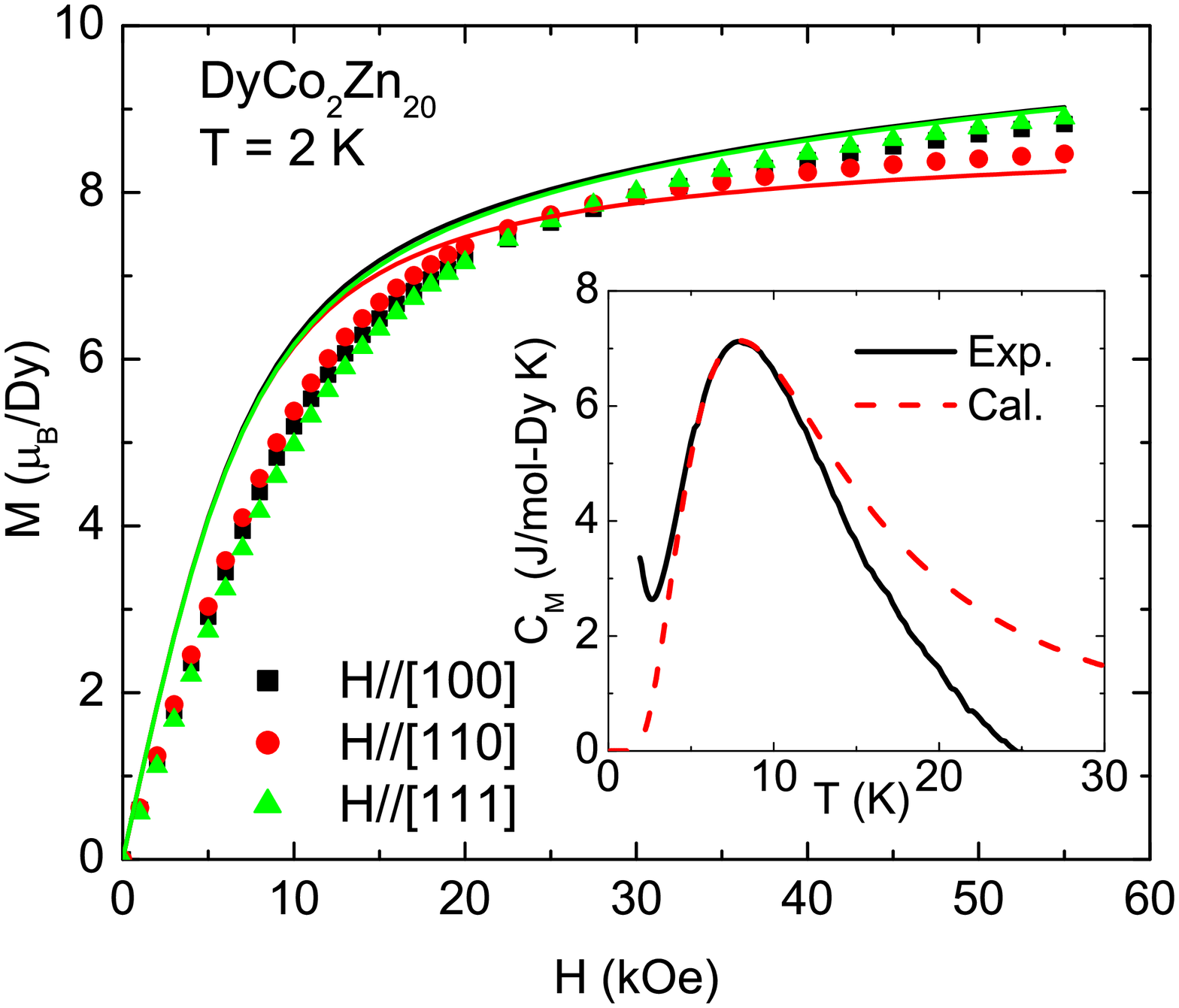}\\
  \caption{(Color online) Field dependent magnetization for DyCo$_2$Zn$_{20}$ along three principle axes. The solid lines present the fitting results. Inset: magnetic part of specific heat. The solid and dashed line present the experimental and calculated result respectively.}
  \label{FigDyCo}
  \end{center}
\end{figure}

\begin{figure}
  \begin{center}
  \includegraphics[clip, width=0.45\textwidth]{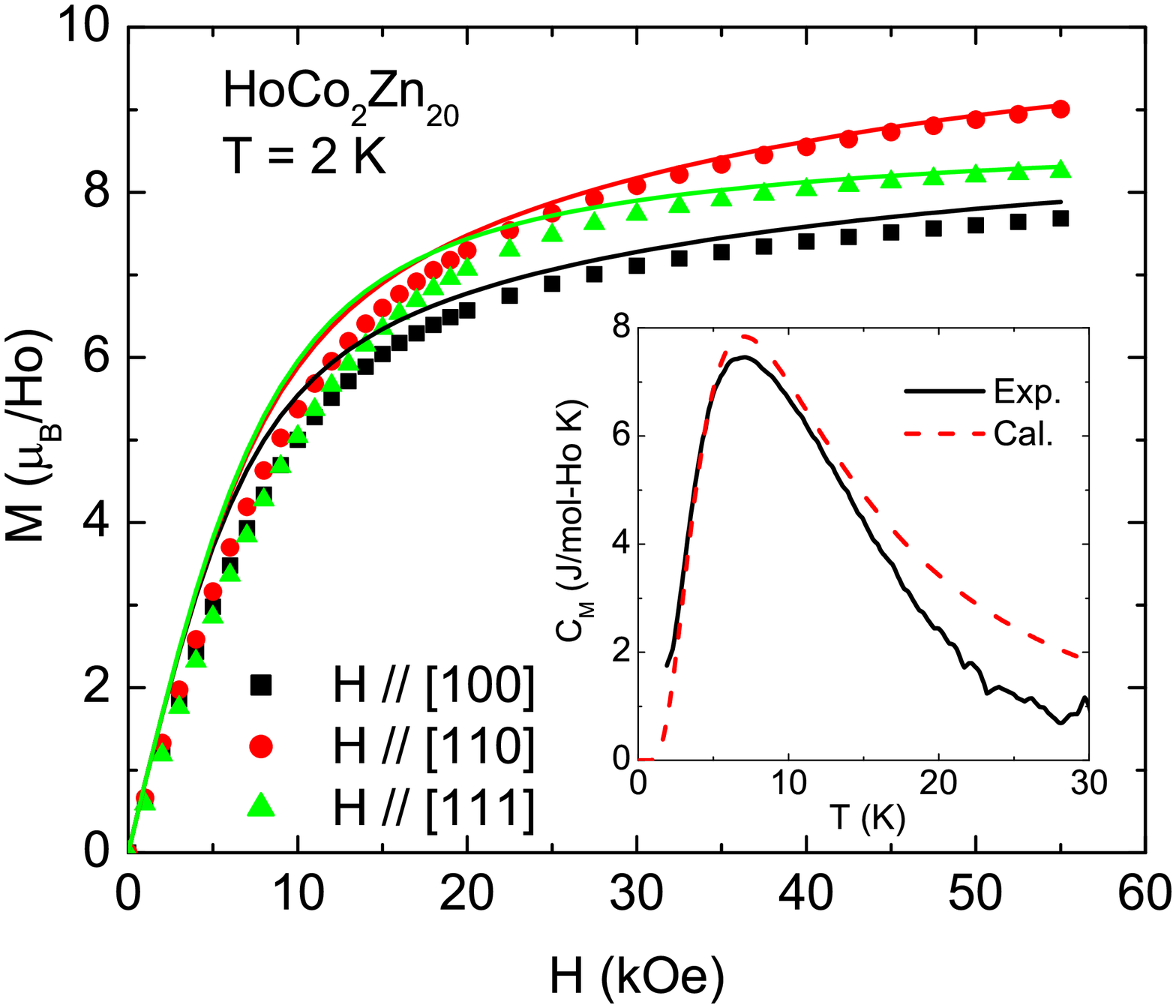}\\
  \caption{(Color online) Field dependent magnetization for HoCo$_2$Zn$_{20}$ along three principle axes. The solid lines present the fitting results. Inset: magnetic part of specific heat. The solid and dashed line present the experimental and calculated result respectively.}
  \label{FigHoCo}
  \end{center}
\end{figure}

\begin{figure}
  \begin{center}
  \includegraphics[clip, width=0.45\textwidth]{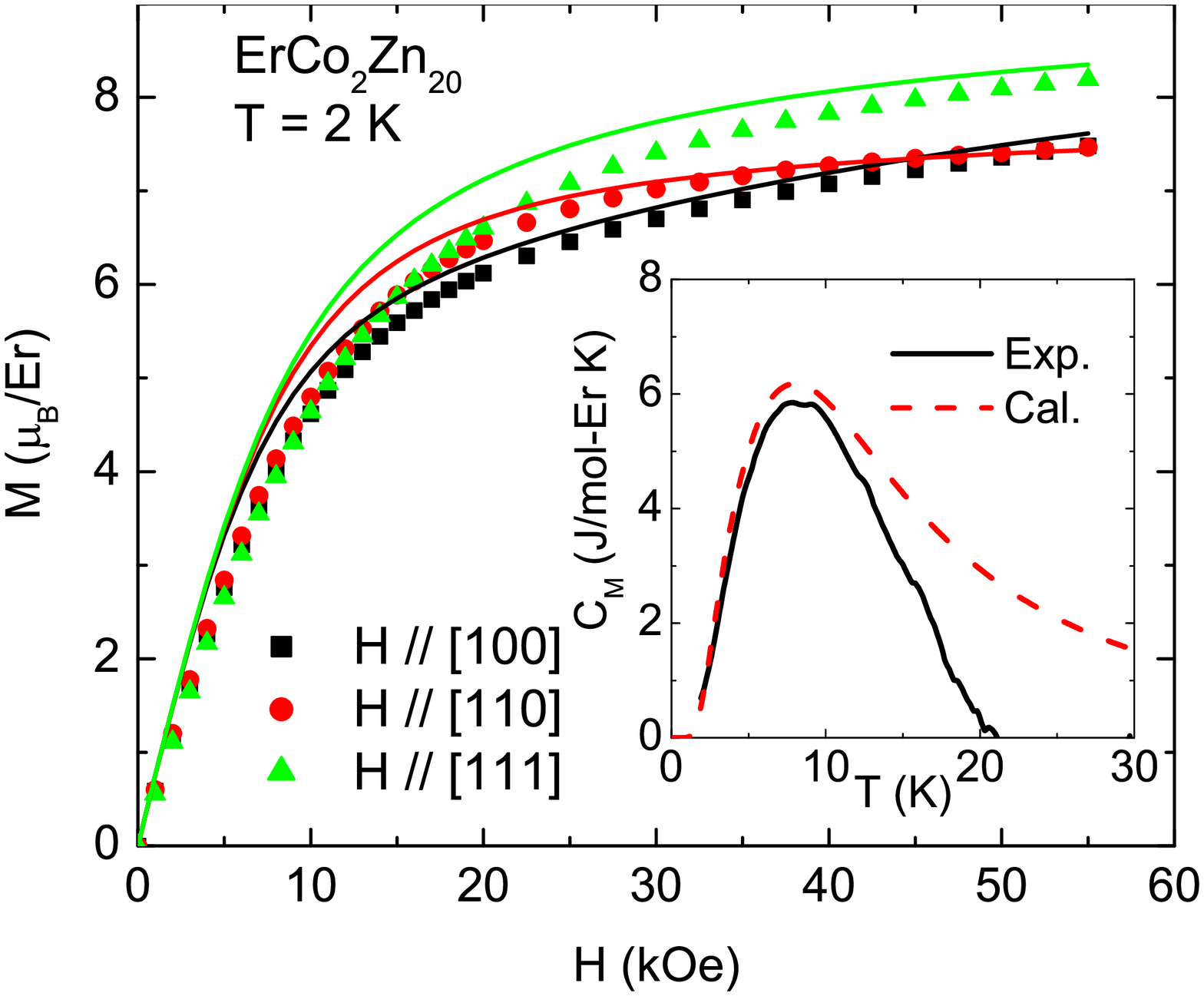}\\
  \caption{(Color online) Field dependent magnetization for ErCo$_2$Zn$_{20}$ along three principle axes. The solid lines present the fitting results. Inset: magnetic part of specific heat. The solid and dashed line present the experimental and calculated result respectively.}
  \label{FigErCo}
  \end{center}
\end{figure}

\begin{figure}
  \begin{center}
  \includegraphics[clip, width=0.45\textwidth]{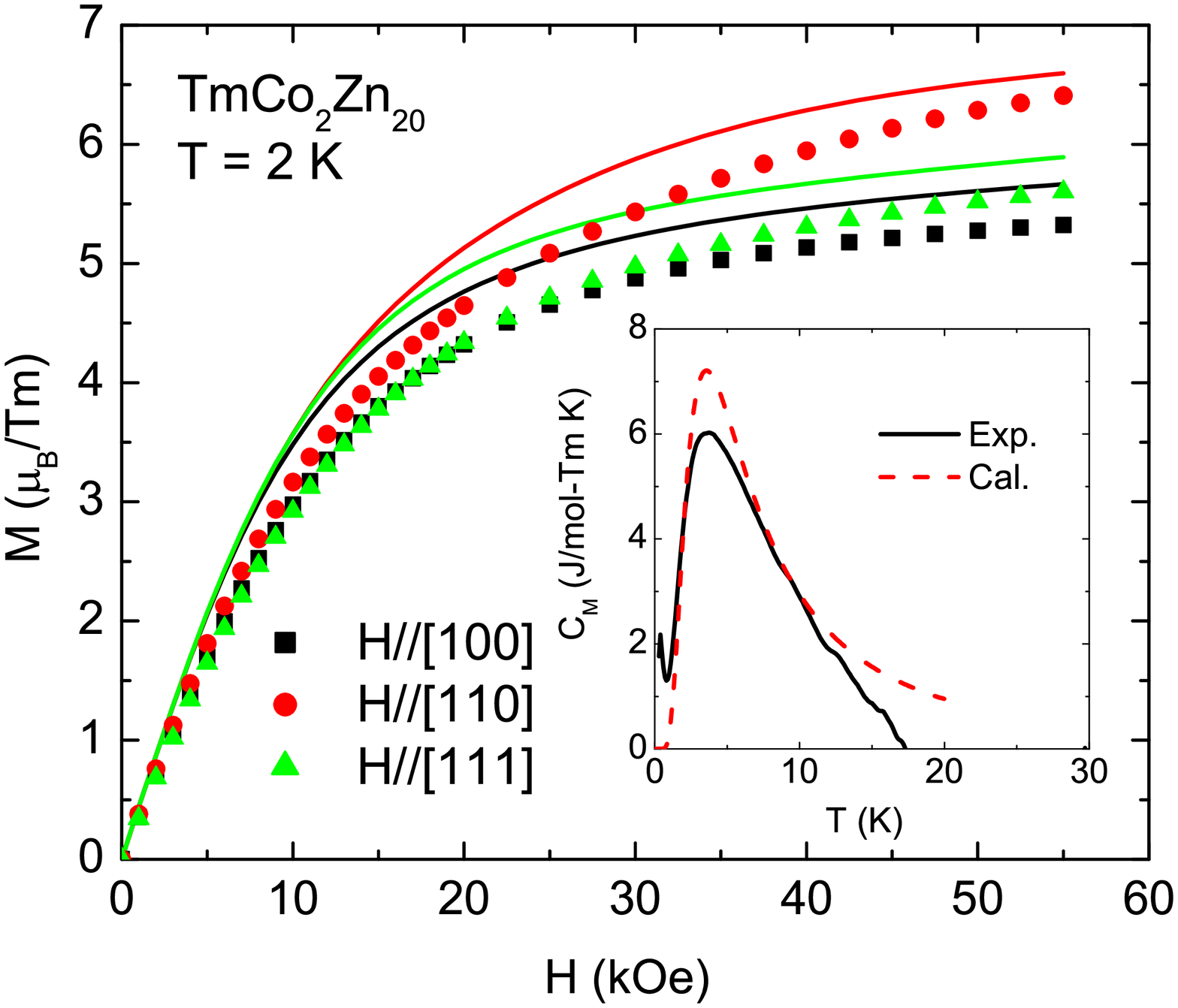}\\
  \caption{(Color online) Field dependent magnetization for TmCo$_2$Zn$_{20}$ along three principle axes. The solid lines present the fitting results. Inset: magnetic part of specific heat. The solid and dashed line present the experimental and calculated result respectively.}
  \label{FigTmCo}
  \end{center}
\end{figure}

Figure \ref{FigTbCo}--\ref{FigTmCo} show data and CEF fitting results of the magnetization at 2~K and the magnetic part of specific heat with the single ion Hamiltonian (ignoring the interaction term) for Tb$_{0.5}$Y$_{0.5}$Co$_2$Zn$_{20}$ and RCo$_2$Zn$_{20}$ (R = Dy - Tm).
The specific heat data for all members are less than the one of YCo$_2$Zn$_{20}$ above 30~K, which is likely due to the errors associated with resolving the difference between the sample's total $C_p$ and the relatively large nonmagnetic contribution. 
Therefore, the fittings of $C_M$ were performed below 20~K.
For R = Dy - Tm, the experimental magnetization data were slightly less than the calculated results.
Such phenomena, more significant for R = Dy and Ho, are most likely due to the still relevant AFM-type of interaction between the local moments.
As shown in table \ref{table3}, the inferred $W$ and $x$ values for all 5 compounds are clustered in a narrow range: $\left| W \right| < 0.1$, $\left| x \right| < 0.25$.
This result, indicating small energy scales of the CEF effect and relatively large $B^0_6$ terms, are roughly consistent with the calculated results based on the point charge model (see Appendix B).
Furthermore, it should be noted that the signs of the $B^0_6$ terms for the calculated results are all consistent with the experimental ones; this is not the case for the $B^0_4$ terms.
This behavior is not difficult to understand, as shown in the Appendix B, the contributions to the CEF splitting are mainly from the CN-16 Frank-Kasper polyhedron formed by 4 NN and 12 NNN Zn neighbors.
For the $B^0_4$ term, the contributions cancel each other by the two sets of neighbors, whereas the contributions for the $B^0_6$ terms is the sum.
Therefore, the $B^0_6$ terms are relatively large and the calculated results are more reliable.

\begin{table}
\caption{\label{table3} Comparison of the CEF parameters of RCo$_2$Zn$_{20}$ compounds (R =Nd, Tb - Yb), determined from magnetization measurements to those calculated in a point charge model. }
\begin{ruledtabular}
\begin{tabular}{llccccccc}
 & & Nd & Tb & Dy & Ho & Er & Tm & Yb\\
\hline
$W$ (K) & exp. & & 0.084 & -0.073 & 0.067 & -0.077 & 0.07 & \\
 & cal. & 0.28 & 0.026 & -0.021 & 0.018 & -0.025 & 0.044 & -0.28\\
\hline
$x$ & exp. & & 0.2 & 0.1 & 0.22 & -0.1 & -0.15 & \\
 & cal. & 0.26 & -0.68 & -0.41 & 0.23 & -0.22 & -0.41 & -0.64\\
\hline
$B^0_4$ ($10^{-4}$ K) & exp. & & 2.8 & -1.2 & 2.5 & 1.3 & 1.75 & \\
 & cal. & 12.2 & -3.0 & 1.4 & 0.7 & -0.9 & -3.0 & 29.6\\
\hline
$B^0_6$ ($10^{-6}$ K) & exp. & & 8.9 & -4.7 & 3.8 & -5 & 7.9 & \\
 & cal. & 81.2 & 1.1 & -0.9 & 1.0 & -1.4 & 3.5 & -81.4\\
\end{tabular}
\end{ruledtabular}
\end{table}
 
\subsection{RFe$_2$Zn$_{20}$ (R = Gd - Tm)}

\begin{figure}
  \begin{center}
  \includegraphics[clip, width=0.45\textwidth]{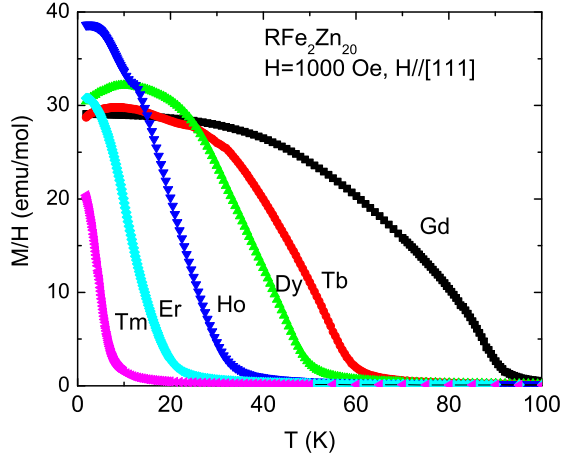}\\
  \caption{(Color online) Temperature dependent magnetization of RFe$_2$Zn$_{20}$ (R = Gd - Tm), divided by applied field $H = 1000$~Oe.}
  \label{FigAllFe1MH}
  \end{center}
\end{figure}

Before discussing each of the compounds in this series separately, an overview of their temperature dependent magnetization data serves as a useful point of orientation.
Figure \ref{FigAllFe1MH} shows $M/H$ versus $T$ (the applied field $H = 1000$ Oe) for R = Gd - Tm members.
In contrast to the Co series compounds, the Fe series compounds all manifest FM ground states with enhanced $T_{\mathrm{C}}$ values, which systematically decrease as R varies from Gd to Tm.
Such enhanced FM ordering has been explained as the result of local moments embedded in the NFFL host, most clearly seen in YFe$_2$Zn$_{20}$ and LuFe$_2$Zn$_{20}$. \cite{jia_nearly_2007} 
This systematic variation of $T_{\mathrm{C}}$ on R is not unexpected for such heavy rare earth compounds when the magnetic interaction between the R ions are associated with the spin part of the Hund's ground state of 4$f$ electrons.

\begin{figure}

  \begin{center}
  \includegraphics[clip, width=0.45\textwidth]{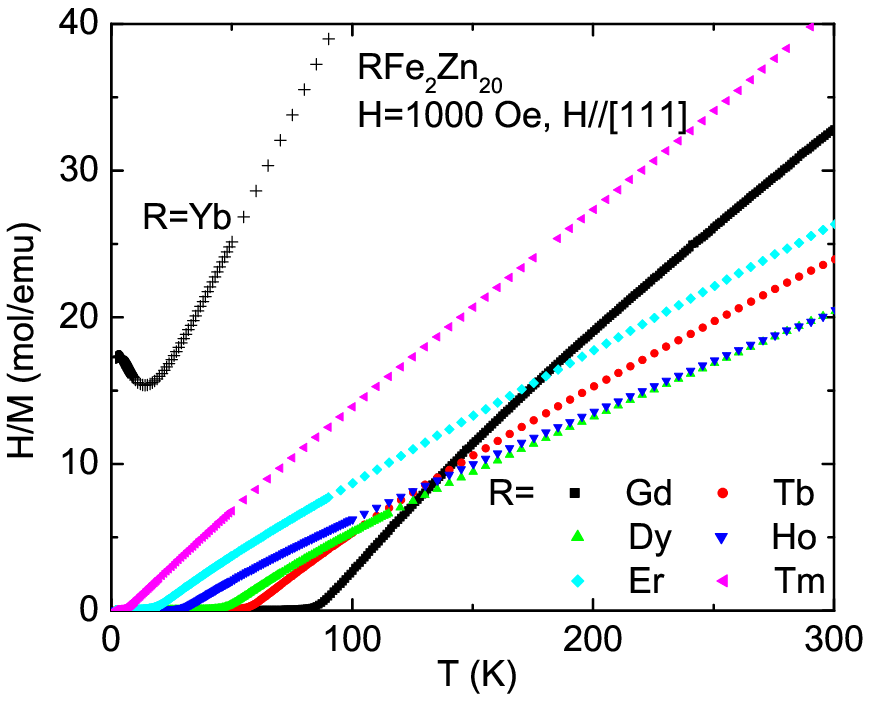}\\
  \caption{(Color online) Applied field ($H = 1000$~Oe) divided by the magnetizations of RFe$_2$Zn$_{20}$ (R = Gd - Tm) as a function of temperature.}
  \label{FigAllFe2HM}
  \end{center}
\end{figure}

The temperature dependent $H/M$ data, approximately equaling inverse susceptibilities [$1/\chi (T)$] in the paramagnetic state, for R = Gd - Tm, as well as for YbFe$_2$Zn$_{20}$, are shown in Fig. \ref{FigAllFe2HM}.
Similar to GdFe$_2$Zn$_{20}$ \cite{jia_nearly_2007}, the $1/\chi (T)$ data sets for R = Tb - Tm follow the CW law [$\chi (T)=C/(T-\theta _C)+\chi _0$] at high temperatures, and deviate from the law when approaching their magnetic ordering.
The effective moment ($\mu _{eff}$) and the paramagnetic Curie temperature ($\theta _C$) data for these 6 compounds are listed in Table \ref{table2}.
All $\mu _{eff}$ values are close to the theoretical value for the Hund's ground state of the trivalent $4f$ electronic configuration.

\begin{table}
\caption{\label{table2} Residual resistivity ratio, $RRR = R(300K)/R(2K)$; paramagnetic Curie temperature, $\theta _C$ (with $\pm 0.5$~K errors) and effective moment, $\mu _{eff}$ (from the CW fit of $\chi (T)$ from 100 K to 300 K, except for GdFe$_2$Zn$_{20}$, which was fitted from 200~K to 375~K \cite{allGd}); Curie temperature, $T_{\mathrm{C}}$; and  saturated moment at 55 kOe along the easy direction, $\mu _{sat}$ for RFe$_2$Zn$_{20}$ compounds (R = Gd - Yb).}
\begin{ruledtabular}
\begin{tabular}{lccccccc}
 & Gd & Tb & Dy & Ho & Er & Tm & Yb \\
\hline
$RRR$ & 8.1 & 7.2 & 15.0 & 10.0 & 13.2 & 10.1 & 31.2\\
$\theta _C$, K & 46 & 30 & 20 & 9 & 0 & -2 & -23\\
$\mu _{eff}$, $\mu _{B}$ & 7.9 & 9.5 & 10.5 & 10.6 & 9.5 & 7.7 & 4.7\\
$T_{\mathrm{C}}$, K & 86 & 58 & 46 & 28 & 17 & 5.5 & \\
$\mu _{sat}$, $\mu _{B}$ & 6.7 & 8.1 & 9.5 & 9.9 & 8.5 & 6.2 & \\
\end{tabular}
\end{ruledtabular}
\end{table}

\subsubsection{TbFe$_2$Zn$_{20}$}

\begin{figure}

  \begin{center}
  \caption{(a) Temperature dependent $M/H$ for TbFe$_2$Zn$_{20}$ ($H = 1000$~Oe); (b) $C_p$; (c) $\rho$  and $\mathrm{d}\rho /\mathrm{d}T$. Upper inset : magnetic part of specific heat. Lower inset: magnetic entropy $S_M$.}
  \includegraphics[clip, width=0.45\textwidth]{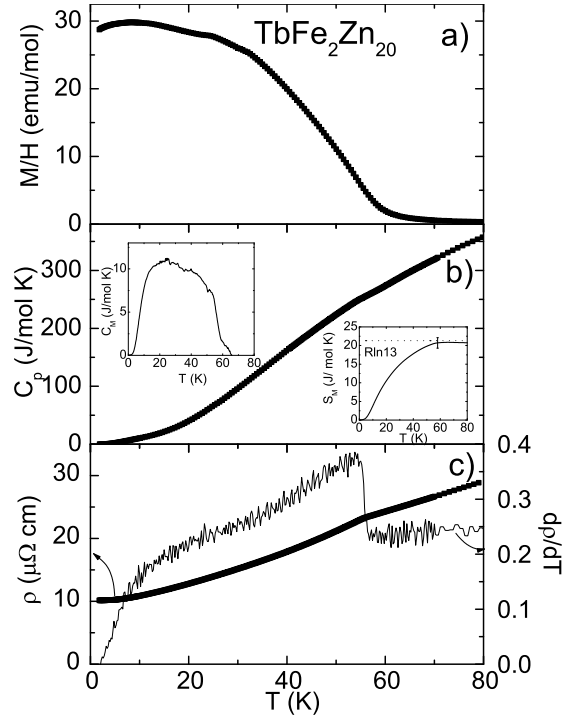}\\
  \label{FigTb1}
  \end{center}
\end{figure}

Temperature dependent $M/H$, specific heat and resistivity data sets for TbFe$_2$Zn$_{20}$ are shown in Fig. \ref{FigTb1}.
The $M(T)/H$ data are consistent with FM order below 60~K, and the magnetic phase transition manifests itself as a faint feature in $C_p$ data, indicating $T_{\mathrm{C}}=56\pm 3$~K.
This is clearer in the magnetic part of specific heat ($C_M$) data (Fig. \ref{FigTb1} upper inset) where $T_{\mathrm{C}}$ is taken as the position of the greateat upward slope.
As shown in the lower inset to Fig. \ref{FigTb1} b, at $T_{\mathrm{C}}$, the magnetic entropy is close to the value for the full degeneracy of the Hund's ground state of Tb$^{3+}$, $R\ln 13$.
As we shall see for the rest of the local moment members (R = Dy - Tm), the released magnetic entropy at $T_{\mathrm{C}}$ for the respective rare earth ion, is close to the full degeneracy value for their Hund's ground state, except for R = Tm.
The $\rho (T)$ data manifests a change in the slope, which could be seen even more clearly on in the $\mathrm{d}\rho /\mathrm{d}T$ data, consistent with a $T_{\mathrm{C}}=56\pm 1$~K.

\begin{figure}

  \begin{center}
  \includegraphics[clip, width=0.45\textwidth]{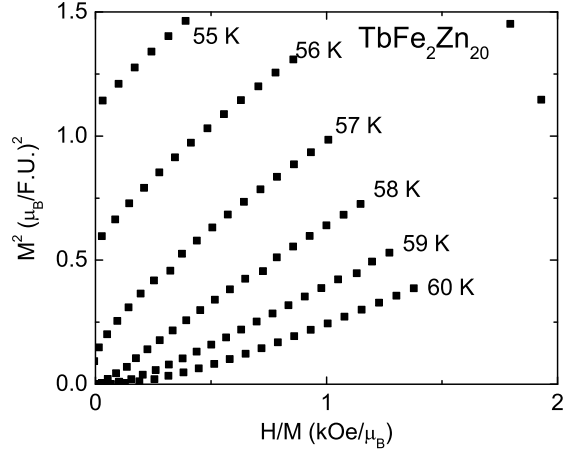}\\
  \caption{Arrott plot of magnetic isothermals for TbFe$_2$Zn$_{20}$.}
  \label{FigTb2}
  \end{center}
\end{figure}

\begin{figure}

  \begin{center}
  \includegraphics[clip, width=0.45\textwidth]{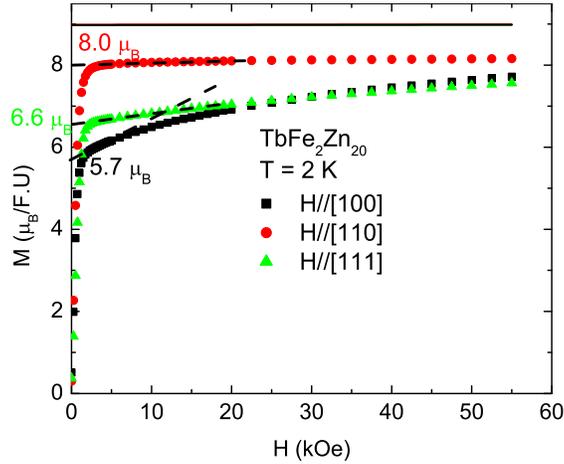}\\
  \caption{(Color online) Field dependent magnetization of TbFe$_2$Zn$_{20}$ along three principle axes at 2~K. The three lines represent the calculated results based on molecular field approximation are all clustered near 9 $\mu _B$ and appear as a single line. The dashed lines and the values present the extrapolate of the magnetization curves and the estimated spontaneous magnetic moments along three directions.}
  \label{FigTb3}
  \end{center}
\end{figure}

Figure \ref{FigTb2} presents a plot of $M^2$ versus $H/M$ (an Arrott plot) isotherms near $T_{\mathrm{C}}$.
The isotherm that most closely goes though the origin is the one closest to $T_{\mathrm{C}}$, giving for this case a value of $58$~K, consistent with the results of the $C_p$ and $\rho (T)$ measurements. 
Figure \ref{FigTb3} shows magnetization versus external field data along 3 different crystallographic directions: [100], [110] and [111], at 2 K.
All of these data sets are consistent with a low temperature FM ground state with moderate anisotropy.
The spontaneous longitudinal magnetic moment in zero applied external field, estimated by the extrapolation of the magnetization curves back to $H=0$, yield $M([110])=8.0 \mu _B$, $M([111])=6.6 \mu _B$, and $M([100])=5.7 \mu _B$.
The ratio of them is very close to $1:\sqrt{2/3}:\sqrt{1/2}$.
Such behavior indicates that the spontaneous magnetic moments along [111] and [100] directions can be understood as the projection of the one along the easy axis, [110].
At 2~K, the saturated moment at 55~kOe along the easy axis, [110], is $ 8.1 \mu _B$, $0.9 \mu _B$ less than the value associated with the Hund's ground state.

\subsubsection{DyFe$_2$Zn$_{20}$}

\begin{figure}

  \begin{center}
  \includegraphics[clip, width=0.45\textwidth]{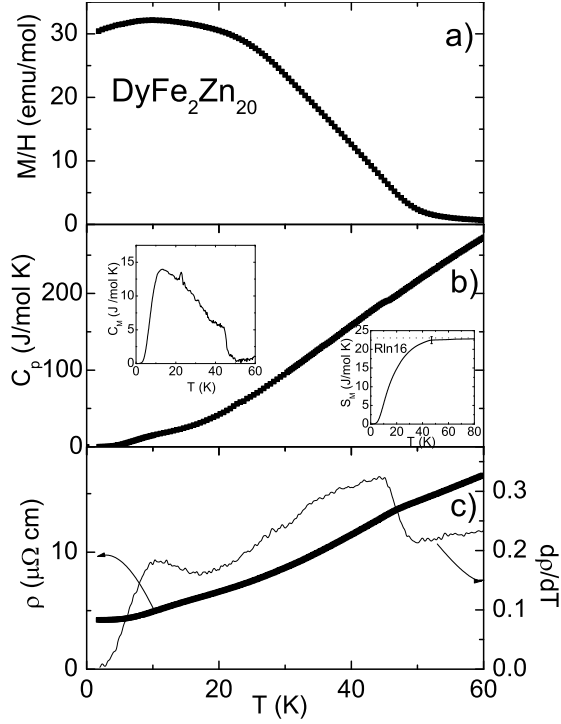}\\
  \caption{(a) Temperature dependent $M/H$ for DyFe$_2$Zn$_{20}$ ($H = 1000$~Oe); (b) $C_p$; (c) $\rho$  and $\mathrm{d}\rho /\mathrm{d}T$. Upper inset: magnetic part of $C_p$. Lower inset: magnetic entropy.}
  \label{FigDy1}
  \end{center}                   
\end{figure}

\begin{figure}

  \begin{center}
  \includegraphics[clip, width=0.45\textwidth]{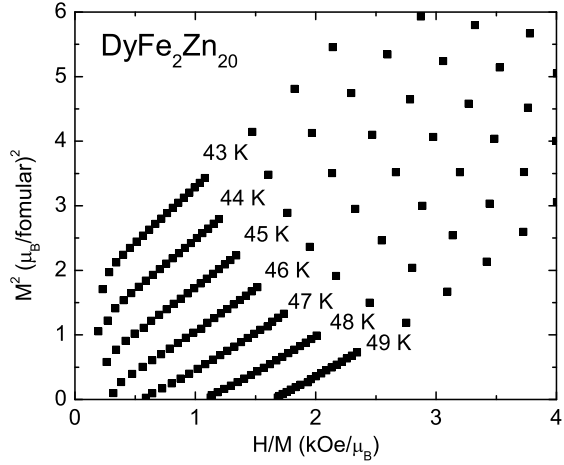}\\
  \caption{Arrott plot of magnetic isothermals for DyFe$_2$Zn$_{20}$.}
  \label{FigDy2}
  \end{center}
\end{figure}

The low field thermodynamic and transport properties of DyFe$_2$Zn$_{20}$ are shown in Fig. \ref{FigDy1}.
The temperature dependent magnetization data (Fig. \ref{FigDy1} a) suggest a FM transition below 50~K. 
The specific heat data show a kink associated with magnetic ordering (Fig. \ref{FigDy1} b), which can be seen more clearly after the subtraction of the non-magnetic background (upper inset) and indicates $T_{\mathrm{C}}=45\pm 1$~K.
This FM transition temperature is further confirmed by a weak change in slope in $\rho (T)$ (associated with the low temperature loss of spin disorder scattering), indicating $T_{\mathrm{C}}=45\pm 2$~K.
Given that the loss of spin disorder scattering in intermetallics often scales with de Gennes parameter \cite{Fournier_transport_1993}, the feature we find in $\rho (T)$ below $T_{\mathrm{C}}$ becomes fainter and fainter as R progresses from Gd to Tm.
These values of $T_{\mathrm{C}}$ are consistent with the result of the Arrott plot analysis, from which a value of $T_{\mathrm{C}}=45\pm 1$~K can be inferred(Fig. \ref{FigDy2}).

It is worth noticing that the specific heat data show a faint shoulder near 10 K, which appears to be a broad peak after the background subtraction, and is coincident with a slope change feature in $\rho (T)$ data.
As seen below, such anomaly below $T_{\mathrm{C}}$ in $C_p$ and $\rho (T)$ data also appears for the members of R = Ho, Er and Tm.
Those anomalies are likely due to the magnetic excitation energy spectrum associated with the Hund's rule multiplet of R$^{3+}$ ions in their FM states. (Further discussion will be presented below.)
 
\begin{figure}
  \begin{center}
  \includegraphics[clip, width=0.45\textwidth]{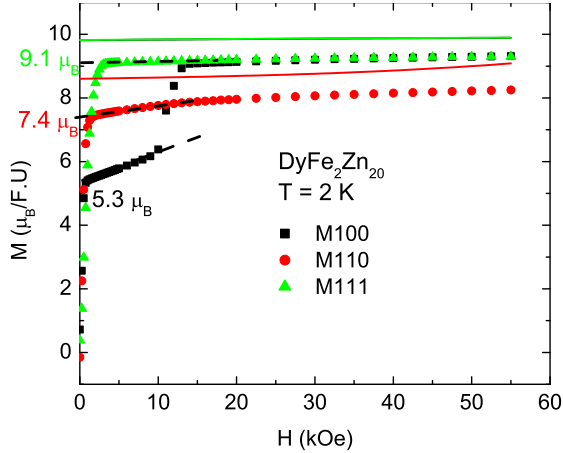}\\
  \caption{(Color online) Field dependent magnetization of DyFe$_2$Zn$_{20}$ at 2~K along three principle axes. The solid lines represent the calculated result based on molecular field approximation (see data analysis part below). The dashed lines and the values present the extrapolate of the magnetization curves and the estimated spontaneous magnetic moments along 3 directions.}
  \label{FigDy3}
  \end{center}
\end{figure}

The 2~K field dependent, magnetization isotherms for DyFe$_2$Zn$_{20}$ are shown in Fig. \ref{FigDy3}.
Compared to TbFe$_2$Zn$_{20}$, the magnetization curves for DyFe$_2$Zn$_{20}$ reveal a slightly more complicated, anisotropic behavior.
The magnetization along [100] direction manifests one metamagnetic phase transition near 12 kOe.
Above this transition, the magnetization along [100] direction is essentially the same as that for the field along the easy [111] axis.
The spontaneous longitudinal magnetization along the three directions, $M([111])=9.1 \mu _B$, $M([110])=7.4 \mu _B$, and $M([100])=5.3 \mu _B$, have a ratio very close to $1:\sqrt{1/2}:\sqrt{1/3}$.
These results indicate that $M([110])$ and $M([100])$ can be seen as the projection of $M([111])$.
The metamagnetic phase transition along [100] can be understood as the process of a classical spin reorientation in a cubic symmetry coordination. 
As in the case for GdFe$_2$Zn$_{20}$ and TbFe$_2$Zn$_{20}$, the saturated moment of DyFe$_2$Zn$_{20}$ at 55~kOe, $9.5 \mu _B$, is slightly less than the value of the Hund's ground state value, $10 \mu _B$.

\subsubsection{HoFe$_2$Zn$_{20}$}

\begin{figure}

  \begin{center}
  \includegraphics[clip, width=0.45\textwidth]{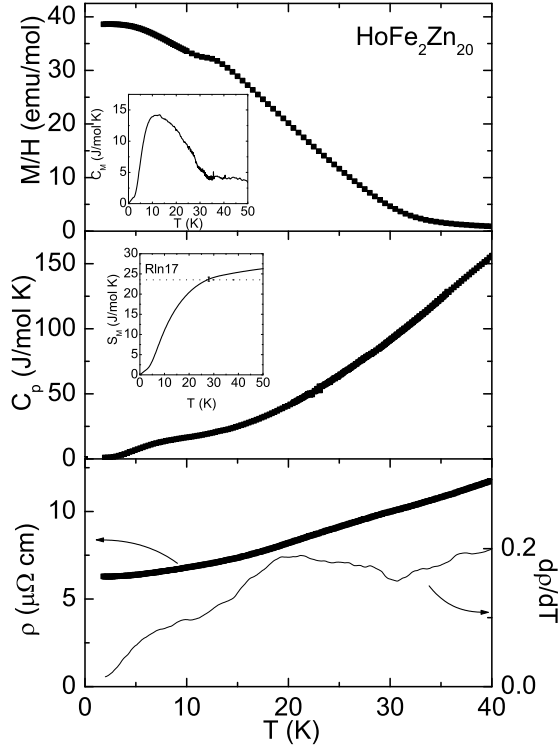}\\
  \caption{(a) Temperature dependent $M/H$ for HoFe$_2$Zn$_{20}$ ($H = 1000$~Oe); (b) $C_p$; (c) $\rho$  and $\mathrm{d}\rho /\mathrm{d}T$. Upper inset: magnetic part of $C_p$ data. Lower inset: magnetic entropy.}
  \label{FigHo1}
  \end{center}
\end{figure}

\begin{figure}

  \begin{center}
  \includegraphics[clip, width=0.45\textwidth]{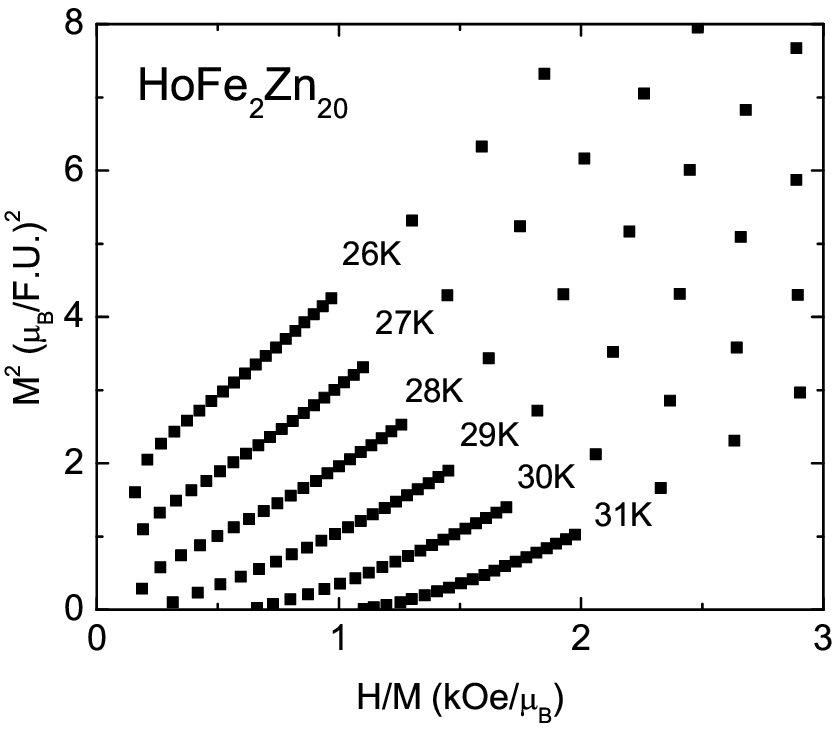}\\
  \caption{Arrott plot of magnetic isothermals for HoFe$_2$Zn$_{20}$.}
  \label{FigHo2}
  \end{center}
\end{figure}

Figure \ref{FigHo1} presents the low field thermodynamic and transport data from measurements on HoFe$_2$Zn$_{20}$.  
The anomalies associated with the FM transition in HoFe$_2$Zn$_{20}$ in the specific heat and resistivity data are relatively weak.
The specific heat anomaly can be associated with $T_{\mathrm{C}} \sim 28$ K, and the $\mathrm{d}\rho /\mathrm{d}T$ data show faint anomaly at this temperature (Fig. \ref{FigHo1}).
The $T_{\mathrm{C}}$ value is determined as $28\pm 1$~K from $C_p$ data, as well as $29\pm 1$~K from $\rho (T)$ data.
This determinate $T_{\mathrm{C}}$ value is consistent with the result of the Arrott plot analysis (Fig. \ref{FigHo2}), which gives $T_{\mathrm{C}}=28\pm 1$~K.

\begin{figure}

  \begin{center}
  \includegraphics[clip, width=0.45\textwidth]{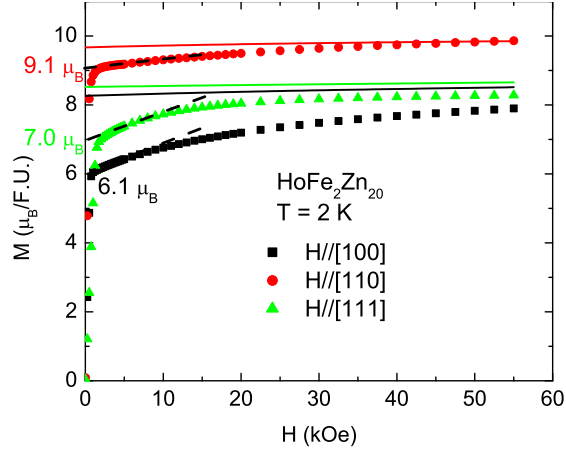}\\
  \caption{(Color online) Field dependent magnetization of HoFe$_2$Zn$_{20}$ at 2~K along three principle axes. The solid lines represent the calculated result based on molecular field approximation (see data analysis part below). The dashed lines and the values present the extrapolate of the magnetization curves and the estimated spontaneous magnetic moments along 3 directions.}
  \label{FigHo3}
  \end{center}
\end{figure}

The low temperature magnetic isotherms for HoFe$_2$Zn$_{20}$ (Fig. \ref{FigHo3}) manifest similar, but obviously larger, anisotropy to the ones for TbFe$_2$Zn$_{20}$.
The ratio of the spontaneous magnetization, $M([110]):M([111]):M([100]) = 9.1 \mu _B:7.0 \mu _B:6.1 \mu _B$ is close to the ratio of $1:\sqrt{2/3}:\sqrt{1/2}$.
This ratio is consistent with the projection of the local moment from the easy [110] axis onto the [111] and [100] axes.
In the external field of 55 kOe, the magnetization along the easy axis, [110], reaches the value of $9.9 \mu _B$, very close to the value of the Hund's ground state, $10 \mu _B$.

\subsubsection{ErFe$_2$Zn$_{20}$}

\begin{figure}

  \begin{center}
  \includegraphics[clip, width=0.45\textwidth]{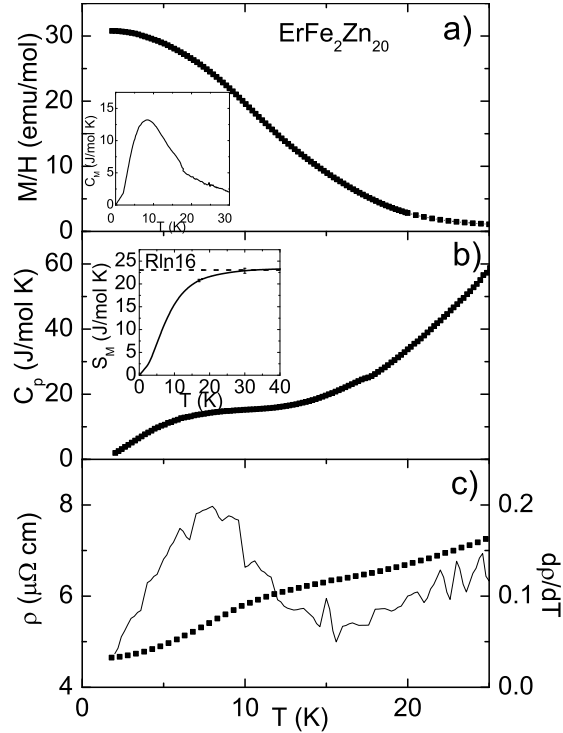}\\
  \caption{(a) Temperature dependent $M/H$ for ErFe$_2$Zn$_{20}$ ($H = 1000$~Oe); (b) $C_p$; (c) $\rho$  and $\mathrm{d}\rho /\mathrm{d}T$. Upper inset: magnetic part of $C_p$ data. Lower inset: magnetic entropy.}
  \label{FigEr1}
  \end{center}
\end{figure}

\begin{figure}

  \begin{center}
  \includegraphics[clip, width=0.45\textwidth]{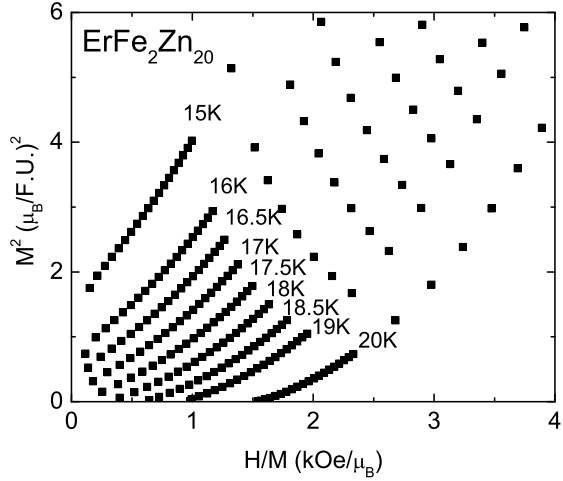}\\
  \caption{Arrott plot of magnetic isothermals for ErFe$_2$Zn$_{20}$.}
  \label{FigEr2}
  \end{center}
\end{figure}

The low field thermodynamic and transport properties of ErFe$_2$Zn$_{20}$ are shown in Fig. \ref{FigEr1}.
The specific heat data show a kink near 18~K (Fig. \ref{FigEr1} b), which can be seen more clearly after the background subtraction (upper inset) and indicates $T_{\mathrm{C}}=18\pm 1$~K.
The resistivity data show no clear anomaly at this temperature.(Fig. \ref{FigEr1})
The released magnetic entropy reaches 21 J/mol~K at $T_{\mathrm{C}}$, 90{\%} of the one associated with the Hund's ground state of Er$^{3+}$, $R\ln 16$ (Fig. \ref{FigEr1} lower inset).
Although $\rho (T)$ data manifest no anomaly at $T_{\mathrm{C}}$, we will see below that the weak anomaly associated with magnetic ordering can be blow up after the background [$\rho (T)$ for LuFe$_2$Zn$_{20}$] subtraction.
The Arrott plot for ErFe$_2$Zn$_{20}$ (Fig. \ref{FigEr2}), although showing non-linear, isothermal curves, demonstrates $T_{\mathrm{C}}=17\pm 0.5$~K with little ambiguity.
The non-linear feature is not unexpected for the $4f$ local moment systems associated with the CEF induced anisotropy. \cite{neumann_arrott_1995}

\begin{figure}
  \begin{center}
  \includegraphics[clip, width=0.45\textwidth]{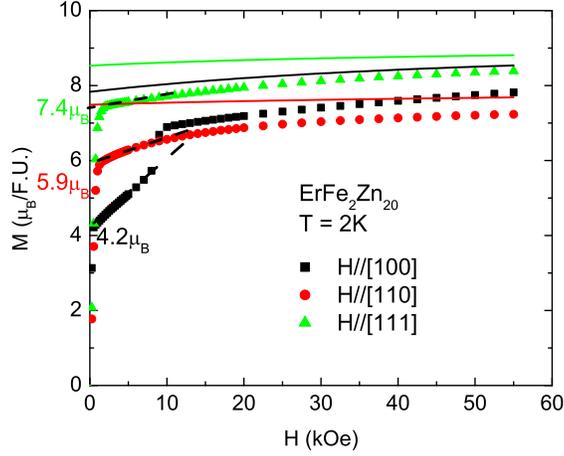}\\
  \caption{(Color online) Field dependent magnetization of ErFe$_2$Zn$_{20}$ at 2~K along three principle axes. The solid lines represent the calculated result based on molecular field approximation (see data analysis part below). The dashed lines and the values present the extrapolate of the magnetization curves and the estimated spontaneous magnetic moments along 3 directions.}
  \label{FigEr3}
  \end{center}
\end{figure}

The magnetic anisotropy of ErFe$_2$Zn$_{20}$ is reminiscent of that of DyFe$_2$Zn$_{20}$: both have the same easy and hard magnetization orientations, [111] and [110] respectively, as well as the metamagnetic transition along the [100] direction (Fig. \ref{FigEr3}).
The ratio of the spontaneous longitudinal magnetic moments, $M([111]):M([110]):M([100]) = 7.4 \mu _B:5.9 \mu _B:4.2 \mu _B$ is also close to the ratio of $1:\sqrt{2/3}:\sqrt{1/3}$.

\subsubsection{TmFe$_2$Zn$_{20}$}

\begin{figure}

  \begin{center}
  \includegraphics[clip, width=0.45\textwidth]{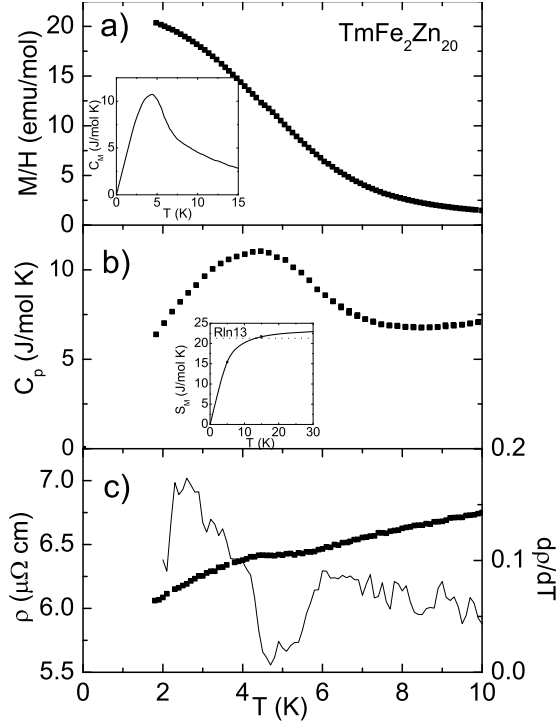}\\
  \caption{(a) Temperature dependent $M/H$ for TmFe$_2$Zn$_{20}$ ($H = 1000$~Oe); (b) $C_p$; (c) $\rho$  and $\mathrm{d}\rho /\mathrm{d}T$. Upper inset: magnetic part of $C_p$ data. Lower inset: magnetic entropy.}
  \label{FigTm1}
  \end{center}
\end{figure}

The low field magnetization, specific heat and resistivity data for TmFe$_2$Zn$_{20}$ are shown in Fig. \ref{FigTm1}.
The temperature dependent magnetization data suggest a FM transition below 10~K (Fig. \ref{FigTm1} a).
However, the specific heat data for TmFe$_2$Zn$_{20}$ only manifest one broad peak at 4.5~K (Fig. \ref{FigTm1} b), which is less like the anomalies associated with $T_{\mathrm{C}}$ for R = Gd - Er, and more like a Schottky anomaly associated with a CEF splitting. 
The resistivity data also show anomaly below 5~K (Fig. \ref{FigTm1} c).
However, at this point, it is difficult to determine whether this anomaly is associated with the magnetic ordering or the CEF splitting of the $4f$ electrons of Tm$^{3+}$ ions. 
As we can see below, after the subtraction of the nonmagnetic background, the anomaly associated with the loss of the spin disorder scattering can be seen more clearly. 

\begin{figure}

  \begin{center}
  \includegraphics[clip, width=0.45\textwidth]{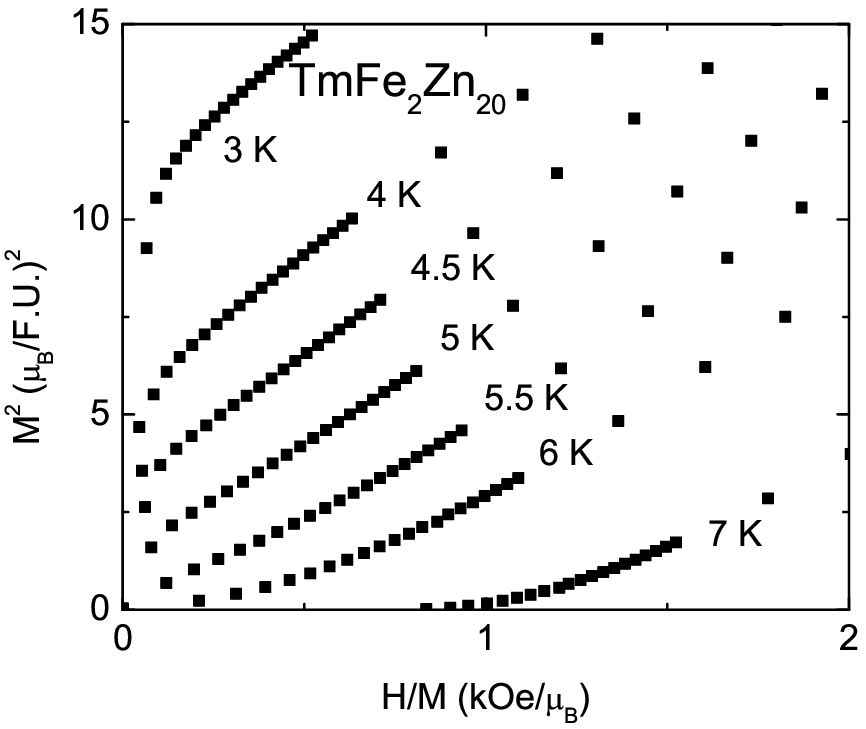}\\
  \caption{Arrott plot of magnetic isothermals for TmFe$_2$Zn$_{20}$.}
  \label{FigTm2}
  \end{center}
\end{figure}

For TmFe$_2$Zn$_{20}$, the Arrott plot analysis provides the reliable criterion for $T_{\mathrm{C}}$ determination.
Figure \ref{FigTm2} shows that $T_{\mathrm{C}}$ can be determined as $5.5\pm 0.5$~K without any ambiguity.
At this temperature, the magnetic entropy is 15 J/mol~K, only 70{\%} of the value of fully released entropy of Hund's ground state of Tm$^{3+}$, $R\ln {13}$ (Fig. \ref{FigTm1} upper inset).

\begin{figure}

  \begin{center}
  \includegraphics[clip, width=0.45\textwidth]{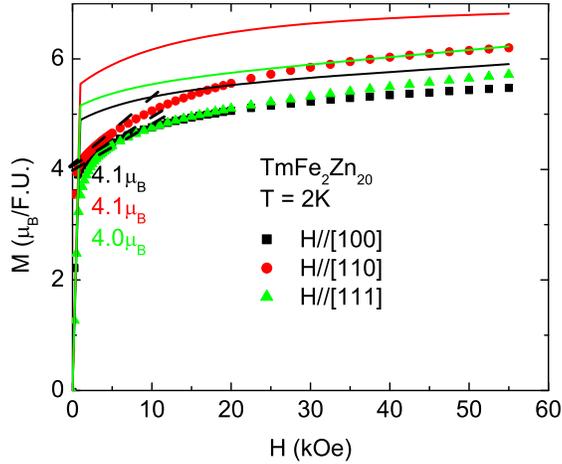}\\
  \caption{(Color online) Field dependent magnetization of TmFe$_2$Zn$_{20}$ at 2~K along three principle axes. The solid lines represent the calculated result based on molecular field approximation (see data analysis part below). The dashed lines and the values present the extrapolate of the magnetization curves and the estimated spontaneous magnetic moments along 3 directions.}
  \label{FigTm3}
  \end{center}
\end{figure}

The low temperature magnetic isotherms for TmFe$_2$Zn$_{20}$ manifest the same easy and hard axis as Tb and Ho members, [110] and [111], respectively (Fig. \ref{FigTm3}).
The spontaneous longitudinal magnetic moments along the three principle axes are all close to $4 \mu _B$.
Such a result may be due to the relatively low value of $T_{\mathrm{C}}$, which makes the spontaneous magnetic moment less anisotropic at 2~K.
The saturated moment along the easy axis reaches $6.2 \mu _B$ at 55~kOe, $0.8 \mu _B$ less than the value of the Hund's ground state, $7 \mu _B$.

\subsection{YbFe$_2$Zn$_{20}$ and YbCo$_2$Zn$_{20}$}

\begin{figure}

  \begin{center}
  \includegraphics[clip, width=0.45\textwidth]{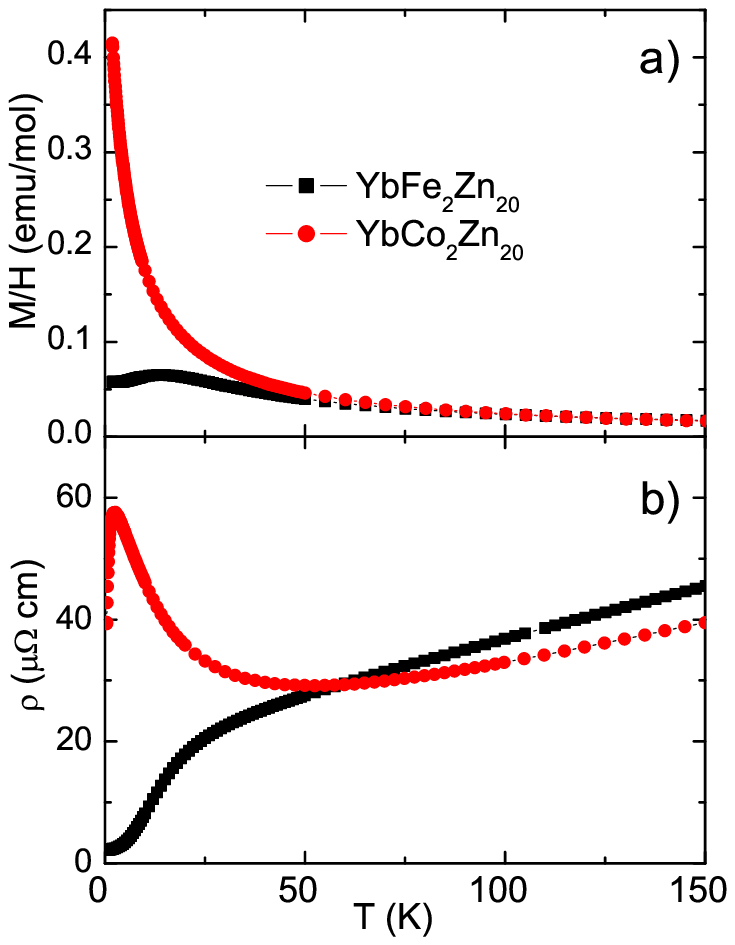}\\
  \caption{Temperature dependent $M/H$ (a) and resistivity (b) for YbFe$_2$Zn$_{20}$ and YbCo$_2$Zn$_{20}$ ($H = 1000$~Oe).}
  \label{FigYb1}
  \end{center}
\end{figure}

The physical properties of YbFe$_2$Zn$_{20}$ and YbCo$_2$Zn$_{20}$ have been discussed in ref. \cite{torikachvili_six_2007}.
They manifest typical heavy Fermion behavior with large electronic specific heat, 520 mJ/mol K$^2$ and 7900 mJ/mol K$^2$, respectively.
Figure \ref{FigYb1} shows temperature dependent susceptibility and resistivity data for these two Yb compounds.
The susceptibility data for YbFe$_2$Zn$_{20}$ manifest a broad, Kondo-type peak about 20~K, indicating a clear loss of local moment behavior, whereas the susceptibility for YbCo$_2$Zn$_{20}$ shows CW behavior down to 1.8 K (Fig. \ref{FigAllCo2HM}), associated with the effective moment value $\mu _{eff} = 4.5 \mu _B$.
Above $\sim 50$~K, $\chi (T)$ for YbFe$_2$Zn$_{20}$ manifests a CW behavior with an effective moment of $4.7 \mu _B$, close to the value of the Hund's ground state of Yb$^{3+}$, $4.5 \mu _B$ (see Fig. \ref{FigAllFe2HM}).
The resistivity data for YbFe$_2$Zn$_{20}$ show a broad shoulder about 30~K, whereas for YbCo$_2$Zn$_{20}$, the resistivity data shows a Kondo resistance minimum about 50~K and a clear coherent peak about 2~K.
These apparently different behaviors for these two Yb-based heavy fermion compounds with virtually identical structures has been explained as the result of significantly different Kondo temperatures: $T_{\mathrm{K}} = 33$~K and 1.5~K for Fe and Co compounds, respectively. 

\section{Data Analysis and Discussion}

\begin{figure}

  \begin{center}
  \includegraphics[clip, width=0.45\textwidth]{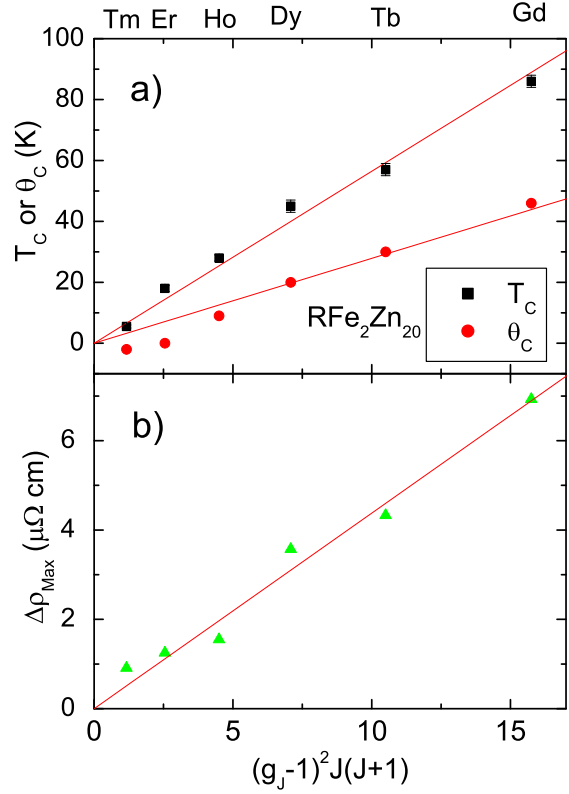}\\
  \caption{(Color online) $T_{\mathrm{C}}$ and $\theta _{\mathrm{C}}$ (a), the maximum value on $\Delta \rho$ (b) with respect to the de Gennes factor for RFe$_2$Zn$_{20}$ (R = Gd - Tm).}
  \label{Figdis1}
  \end{center}
\end{figure}

As shown in Fig. \ref{Figdis1} a, the $T_{\mathrm{C}}$ values of RFe$_2$Zn$_{20}$ compounds (R = Gd - Tm) scale fairly well with the de Gennes factor, $dG=(g_J-1)^2J(J+1)$, consistent with a RKKY interaction.
All of the $\theta _C$ values for each compounds are smaller than their respective $T_{\mathrm{C}}$ values (for R = Er and Tm, the values of $\theta _C$ are even negative).
These small $\theta _C$ values are consistent with the deviation of $\chi (T)$ from the CW law (Fig. \ref{FigAllFe2HM}).
As observed in the case of pseudo-ternary compounds Gd$_x$Y$_{1-x}$Fe$_2$Zn$_{20}$, such deviation can be explained as a result of increasing coupling between the local moments embedded in the strongly temperature dependent, polarizable matrix, YFe$_2$Zn$_{20}$ or LuFe$_2$Zn$_{20}$. \cite{jia_nearly_2007}

Previous studies show that the magnetization of GdFe$_2$Zn$_{20}$ at base temperature is nearly isotropic with a slightly reduced saturated moment ($\sim 0.5\mu _B$ less than the value of the Hund's rule ground state of Gd$^{3+}$). 
For R = Tb - Tm, the magnetization anisotropy at base temperature is significant, and correlates with the easy and hard axes of the respective RCo$_2$Zn$_{20}$ analogues.
Such behavior indicates the anisotropy of the RFe$_2$Zn$_{20}$ (R = Tb - Tm) compounds may mainly be due to the CEF effect on the R$^{3+}$ ions.
The $M(H)$ curves at 2~K manifest divided behavior with R =  Tb, Ho and Tm on one hand and R = Dy and Er on the other hand: for R = Tb, Ho and Tm, the magnetization processes are gradual along all 3 principal axes; for R = Dy and Er, the magnetization data along [100] direction shows clear and sharp metamagnetic transition.
Both types of magnetization processes (gradual increase and metamagnetic transition) are common for the FM ordered $4f$ local moments with CEF anisotropy associated with the R in a cubic point symmetry, and can be understood in terms of the purification of the CEF split $4f$ electronic wave function due to the Zeeman effect of the external field, and the rotation of the local moment. \cite{magnetism_metal_alloy}
Given that Tb$^{3+}$ and Tm$^{3+}$, as well as Dy$^{3+}$ and Er$^{3+}$ ions have same total $4f$ electronic Hund's rule ground state quantum number ($\textbf{J}=6$ and $15/2$ respectively), the similar magnetic anisotropy indicates similar CEF effect for the two sets of rare earth ions, respectively.
       
In order to better understand the magnetic anisotropy of RFe$_2$Zn$_{20}$ compounds (R = Tb - Tm), the CEF effect acting on the R ions must to be considered.
However, multiple difficulties associated with the strongly polarizable back ground [Y(Lu)Fe$_2$Zn$_{20}$] as well as the strong magnetic interaction, make the determination of the CEF parameters hard.
For example, in order to reduce the magnetic interaction, the magnetic R$^{3+}$ ions were placed into a dilute coordination, R$_x$Y$_{1-x}$Fe$_2$Zn$_{20}$ or R$_x$Lu$_{1-x}$Fe$_2$Zn$_{20}$.
A FM ground state has been found even for very dilute magnetic R concentration: it was found that Tb$_{0.05}$Y$_{0.95}$Fe$_2$Zn$_{20}$, Dy$_{0.05}$Y$_{0.95}$Fe$_2$Zn$_{20}$ as well as Ho$_{0.1}$Y$_{0.9}$Fe$_2$Zn$_{20}$ manifest FM ordering above 2~K.
For such small $x$, the background subtraction (magnetization and/or specific heat of YFe$_2$Zn$_{20}$ or LuFe$_2$Zn$_{20}$), as well as the uncertainty of $x$, make the fitting process unreliable.

\begin{figure}
  \begin{center}
  \includegraphics[clip, width=0.45\textwidth]{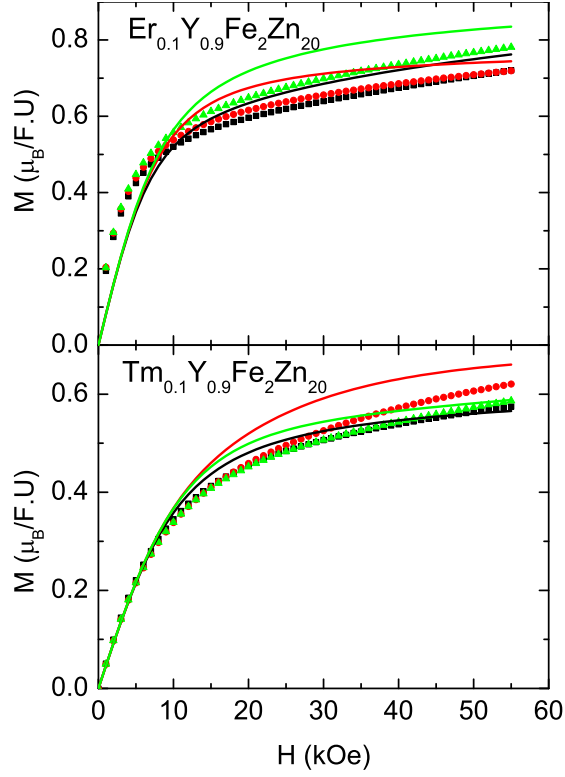}\\
  \caption{(Color online) Field dependent magnetization for Er$_{0.1}$Y$_{0.1}$Fe$_2$Zn$_{20}$ (a) and Tm$_{0.1}$Y$_{0.1}$Fe$_2$Zn$_{20}$ (b) along three principle axes at 1.85~K. The solid lines present the calculated magnetization by using the single-ion Hamiltonian and the CEF coefficients determined from the respective Co members.}
  \label{FigErTmFe}
  \end{center}
\end{figure}

On the other hand, due to the very similar R coordination and the lattice parameters for Fe and Co series, the CEF parameters determined from RCo$_2$Zn$_{20}$ compounds should be close to those for the RFe$_2$Zn$_{20}$ compounds, for respective R members.
Figure \ref{FigErTmFe} shows that the anisotropic magnetization isotherms for the pseudo-ternary Fe compounds, Er$_{0.1}$Y$_{0.9}$Fe$_2$Zn$_{20}$ and Tm$_{0.1}$Y$_{0.9}$Fe$_2$Zn$_{20}$, which still manifest a paramagnetic state at the base temperature, are close to the calculated results from the determined CEF parameters of the respective Co compounds.
The calculated results also fairly well mimic the crossing behavior of the magnetization along [110] and [100] directions for R = Er, as well as along the [111] and [100] directions for R = Tm.
For all three directions, the calculated results are slightly larger than the experimental ones, which is most likely due to the $\pm 0.02$ uncertainty of the nominal $x$ value.
The larger magnetization for Er$_{0.1}$Y$_{0.9}$Fe$_2$Zn$_{20}$ than the calculated results below 10~kOe is consistent with residual FM interactions between the Er$^{3+}$ local moments.

The magnetization along the three axes for the all Fe compounds were calculated based on the molecular field approximation in a self-consistent manner.
In the single-ion Hamiltonian for the R$^{3+}$ ions (Eqn. \ref{eqn:3}), with the molecular field approximation, the magnetic interaction term is written as: 
\begin{equation}
\mathcal{H}_{exc} = g_J \mu _B\vec{J}\cdot \vec{H}_M,
\label{eqn:6}
\end{equation}
where $H_M$ is the molecular field. It obeys the self-consistent condition:
\begin{equation}
H_M = \lambda g\mu _B \left\langle\vec{J}\right\rangle,
\label{eqn:7}
\end{equation}
\begin{equation}
\left\langle\vec{J}\right\rangle = \frac{\sum_0 J_n \exp{(-E_n/k_BT)}}{\sum_0\exp{(-E_n/k_BT)}},
\label{eqn:8}
\end{equation}
where $J_n$ and $E_n$ are the eigenvalues and eigenenergies of the nth eigenfunction; $\lambda $ is the molecular field constant which can be obtained from the ordering temperature: $\lambda = \frac{3k_BT_{\mathrm{C}}}{\mu ^2_{eff}}$.

The calculated magnetizations were compared with the experimental results in Figs. \ref{FigTb3}, \ref{FigDy3}, \ref{FigHo3}, \ref{FigEr3} and \ref{FigTm3}.
All these calculated magnetization values are obviously larger than the experimental results.
This difference is probably due to (i) the molecular field approximation over-estimating the molecular field constant as well as the internal field, and (ii) the induced moments from the Fe site aligning in an antiparallel manner with respect to the R$^{3+}$ local moments (as in the case of GdFe$_2$Zn$_{20}$ \cite{jia_nearly_2007}). 

\begin{figure}
  \begin{center}
  \includegraphics[clip, width=0.45\textwidth]{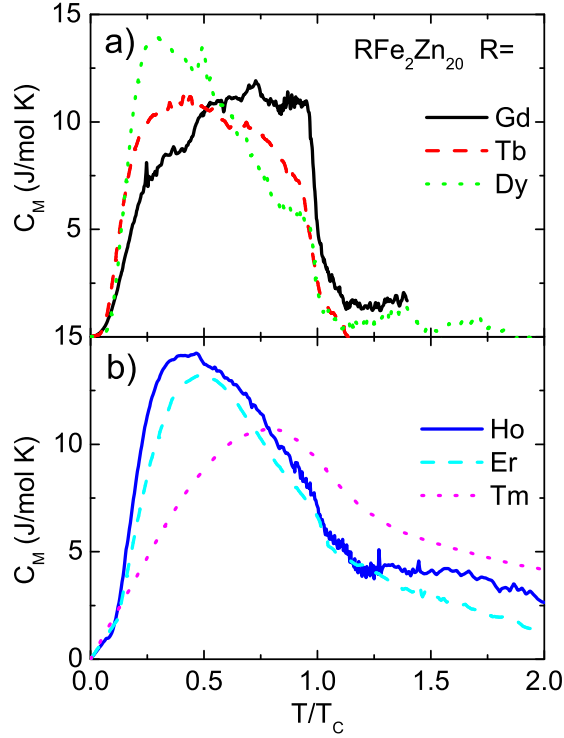}\\
  \caption{(Color online) Magnetic part of specific heat versus $T/T_{\mathrm{C}}$ for RFe$_2$Zn$_{20}$ (R = Gd - Tm).}
  \label{Figdis2}
  \end{center}
\end{figure}

Figure \ref{Figdis2} shows the magnetic part of specific heat as a function of $T/T_{\mathrm{C}}$ for RFe$_2$Zn$_{20}$ (R = Gd - Tm).
The magnetic ordering temperature ($T_{\mathrm{C}}$) of R = Gd - Er members manifests itself as the position of maximum slope, with a decreasing sharpness as R varies from Gd to Er.
TmFe$_2$Zn$_{20}$ does not appear to have any anomaly in the $C_M$ data at $T_{\mathrm{C}}$.
Below $T_{\mathrm{C}}$, the data sets for R = Dy - Tm show a broad peak, which shifts closer to its $T_{\mathrm{C}}$ as R varies from Dy to Tm, whereas the data for $\mathrm{GdFe_2Zn_{20}}$ show no explicit peak.
If the broad peaks are corresponding to the magnetic excitation energy spectrum associated with CEF effect, then the relative positions of these peaks to $T_{\mathrm{C}}$, to some extend, indicate the ratio of the energy scales of the CEF splitting (for a single ion) to the magnetic interaction.
The shift of the peak position as R varies from Dy to Tm indicates that the energy scale of the magnetic order decreases relative to the CEF splitting.
Such phenomena is consistent with the analysis on the magnetic part of entropy: as shown in the insets of Figs. \ref{FigTb1}, \ref{FigDy1}, \ref{FigHo1}, \ref{FigEr1} and \ref{FigTm1}, Tb, Dy and Ho compounds manifest fully released $S_M$ at their $T_{\mathrm{C}}$; whereas Er and Tm compounds still release part of $S_M$ above their $T_{\mathrm{C}}$, which indicates that,  unlike R = Gd - Ho members, the CEF splitting for the $4f$ electronic configuration of the Tm$^{3+}$ and Er$^{3+}$ may extend above magnetic ordering temperature.

\begin{figure}
  \begin{center}
  \includegraphics[clip, width=0.45\textwidth]{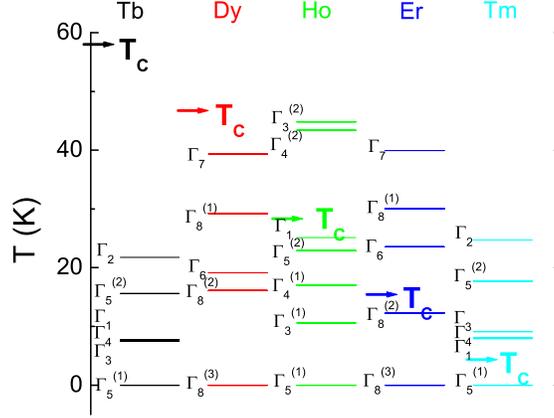}\\
  \caption{(Color online) Single-ion CEF splitting energy levels for RCo$_2$Zn$_{20}$ (R = Tb - Tm). The arrows present the $T_{\mathrm{C}}$ values for RFe$_2$Zn$_{20}$ with respective R.}
  \label{Figlevels}
  \end{center}
\end{figure}

Based on the assumption that the Fe and Co series have similar CEF splitting (for a similar R ions), the comparison between the magnetic ordering temperature and the CEF splitting for different R ions is qualitatively diagrammatized in Fig. \ref{Figlevels}.
The levels represent the single ion, CEF splitting of the Hund's ground state of $4f$ electronic configuration of R$^{3+}$, determined from RCo$_2$Zn$_{20}$ and the arrows represent the $T_{\mathrm{C}}$ values of RFe$_2$Zn$_{20}$.  
The $T_{\mathrm{C}}$ value is comparable with the highest energy level of CEF splitting for R = Ho.
For R = Er and Tm, the $T_{\mathrm{C}}$ values is about $\frac{1}{2}$ and $\frac{1}{5}$ of the highest CEF levels, respectively. 
This diagram, though it cannot be used to determine the precise energy splitting of the RFe$_2$Zn$_{20}$ compounds (the CEF levels have been strongly modulated and mixed by the interaction energy), is qualitatively consistent with the specific heat measurements, and indicates that, at least for TmFe$_2$Zn$_{20}$, the CEF energy splitting already happens well above its $T_{\mathrm{C}}$.
In summary, is appears plausible that, due to extremely similar liganal environments, equivalent members of the RFe$_2$Zn$_{20}$ and RCo$_2$Zn$_{20}$ series have similar CEF splitting schemes.

\begin{figure}
  \begin{center}
  \includegraphics[clip, width=0.45\textwidth]{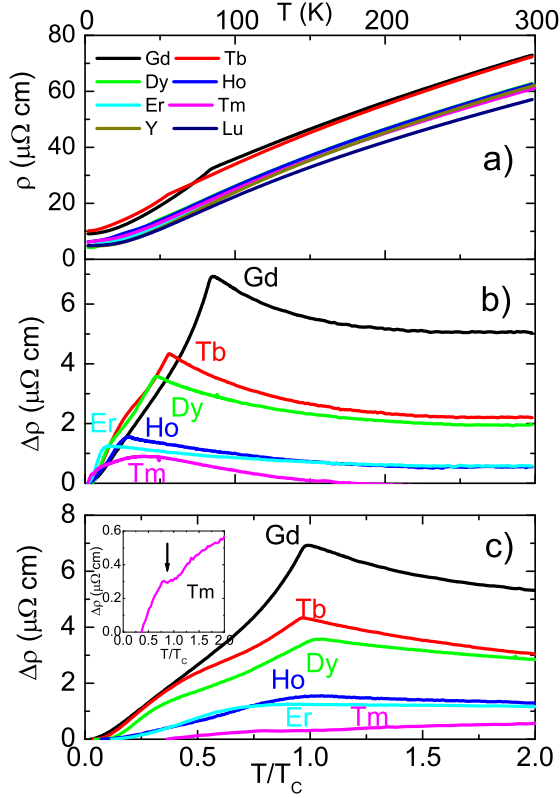}\\
  \caption{(Color online) (a):  $\rho $ versus $T$, (b):  $\Delta \rho $ versus $T$, (c): $\Delta \rho $ versus $T/T_{\mathrm{C}}$ for RFe$_2$Zn$_{20}$ (R = Gd - Tm). Inset: the blow up $\Delta \rho $ data for TmFe$_2$Zn$_{20}$. The arrow presents the FM ordering temperature.}
  \label{Figdis3}
  \end{center}
\end{figure}

Further insight can also be gained from a careful revisiting of the transport data.
The total resistivity of RFe$_2$Zn$_{20}$ (R = Gd - Yb) can be written as:
\begin{equation}
\rho (T)=\rho _{0}+\rho _{ph}(T)+\rho _{mag}(T),
\label{eqn:9}
\end{equation}
where $\rho _{mag}$ is associated with scattering from the $4f$ moments and the spin fluctuation of itinerant electrons.
As seen in Fig. \ref{Figdis3} a, for the whole series above 250~K, the resistivity data sets show essentially linear behavior with slopes differing by less than 12\%, within the estimated dimension error ($\pm 10${\%}) of these bar-like-shape samples.
These similar, high temperature behaviors indicate that, in the high temperature limit, the magnetic scattering is saturated, and the phonon scattering is essentially similar for the whole series (due to the very dilute nature of the R ions).
Therefore, the magnetic contribution to the resistivity can be estimated by (1) subtracting residue resistivity, $\rho _0$ (2) normalizing the high temperature slope of all $\rho (T)$ to that of LuFe$_2$Zn$_{20}$ and then (3) subtracting the $\rho _{Lu}(T)-\rho _{Lu0}$ data from the normalized data. The result is written as:

\begin{equation}
\Delta \rho (T)=(\rho _{\mathrm{R}}-\rho _{\mathrm{R}0})\frac{\frac{\mathrm{d}\rho _{|mathrm{R}}}{\mathrm{d}T}\mid _{275\mathrm{K}}}{\frac{\mathrm{d}\rho _{\mathrm{Lu}}}{\mathrm{d}\mathrm{T}}\mid _{275\mathrm{K}}}-(\rho _{\mathrm{Lu}}-\rho _{\mathrm{Lu}0}).
\label{eqn:10}
\end{equation}
 
As shown before, the subtraction background $\rho _{Lu}(T)$ already includes the scattering associated with the spin fluctuation of itinerant electrons.
Thus, $\Delta \rho $ will not only include the scattering from the $4f$ moments, but will also include scattering associated with the interaction between the $4f$ moments and itinerant electrons.
Figure \ref{Figdis3} b and c show $\Delta \rho $ versus temperature, as well as normalized temperature ($T/T_{\mathrm{C}}$) for R = Gd - Tm.
For R = Gd - Er, a pronounced upward cusp, whose height decreases from Gd to Er, is centered about $T_{\mathrm{C}}$, whereas TmFe$_2$Zn$_{20}$ manifests a broad feature and only very weak anomaly around its $T_{\mathrm{C}}$ (see the blow-up inset of Fig. \ref{Figdis3}).
As shown in Fig. \ref{Figdis1} b, the maximum values of the cusps for different R scale with the de Gennes factor, which indicates that the decrease of $\Delta \rho $ with $T$ below $T_{\mathrm{C}}$ is the result of a loss of spin disorder scattering of conduction electrons, associated with the $4f$ local moment.
However, as found in the pseudo-ternary compounds Gd$_x$Y$_{1-x}$Fe$_2$Zn$_{20}$,\cite{jia_GdY_2007} the decreasing behavior of $\Delta \rho $ with increasing $T$ above $T_{\mathrm{C}}$ is more conspicuous and must come from a different conduction electron scattering process (simple models of $\rho (T)$ due to magnetic scattering cannot explain this anomaly \cite{craig_transport_1967, fisher_resistive_1968}).
Giving that RT$_2$Zn$_{20}$ compounds only manifest this behavior when the local moments are embedded in the highly polarizable background (GdCo$_2$Zn$_{20}$ does not show this behavior \cite{jia_nearly_2007}), this anomaly is thought to be associated with the spin fluctuation of the $3d$ electrons.
Also appearing in the resistivity of RCo$_2$ (R = Gd - Tm)\cite{gratz_transport_1995}, this decreasing behavior of $\Delta \rho $ with increasing $T$ above $T_{\mathrm{C}}$ has been explained as the result of the increase of the spin fluctuation of $3d$ electrons, which is provided by the increasing, nonuniform fluctuating $4f$-$d$ electron exchange interaction, as the temperature approaches $T_{\mathrm{C}}$ in the paramagnetic state.
Since both Y(Lu)Co$_2$ and Y(Lu)Fe$_2$Zn$_{20}$ are classical examples of NFFLs, such an anomaly could be associated with these strongly correlated electron systems.
On the other hand, considering that the Hund's ground state of Tm$^{3+}$ has been significantly split above FM ordering for TmFe$_2$Zn$_{20}$, it is not unexpected that the conduction electron scattering process manifests a different behavior associated with the CEF effect.

Finally, the nearly FM compounds: YFe$_2$Zn$_{20}$ and LuFe$_2$Zn$_{20}$ also merit some further discussion.
Shown in Fig. \ref{FigYLu1M}, the low field susceptibility (H = 10 kOe) manifests a maximum near 6~K and 8~K for YFe$_2$Zn$_{20}$ and LuFe$_2$Zn$_{20}$ respectively.
Such a maximum in the temperature dependent susceptibility also appears for other examples of nearly FM compounds.
For example, Pd manifests $T_{max} \sim 70$~K \cite{pd_kia}; YCo$_2$ and LuCo$_2$ manifests $T_{max} \sim 100$~K\cite{yco2_kia}; and TiBe$_2$ manifests $T_{max} \sim 10$~K\cite{tibe2_kia}.
Another interesting phenomena in nearly FM materials is the so-called itinerant electron metamagnetism (IEM), which is an applied magnetic field induced, first order, phase transition between a paramagnetic state and spin polarized state \cite{IEM}.
Experimentally, IEM has been observed for YCo$_2$ and LuCo$_2$ around 70 T. \cite{goto_itinerant_1989, goto_IEMluco2_1990}  
Within the framework of Landau theory, the maximum in temperature dependent susceptibility is thought to be related to IEM.\cite{shimizu_IEM_1981} 
The magnetic part of the free energy $\Delta F$ can be writen as the function of the magnetic moment $M$:
\begin{equation}
\Delta F = \frac{1}{2}aM^2+\frac{1}{4}bM^4+\frac{1}{6}cM^6~,
\label{eqn:11}
\end{equation}
where $a$, $b$ and $c$ are the Landau expansion coefficients. 

As shown by Shimizu \cite{shimizu_IEM_1981}, the condition for the existence of IEM is:
$a > 0$, $b < 0$, $c > 0$ and $\frac{3}{16} < \frac{ac}{b^2} < \frac{9}{20}$.
Within the framework of the spin fluctuation theory, Yamada \cite{yamada_1993} generalized this work by introducing a temperature dependent function of the mean square amplitude of spin fluctuations.
These theoretical works demonstrated that the existence of IEM is associated with the maximum in $\chi (T)$ by means of the factor of $\frac{ac}{b^2}$, which can be estimated as:
\begin{equation}
\frac{ac}{b^2}=[1-\frac{\chi (0)}{\chi (T_{max})}]^{-1}.
\label{eqn:12}
\end{equation}
Furthermore, the IEM can only happen below $T_{max}$.
These results seem to be consistent with the experimental results in various itinerant electronic systems. \cite {goto_itinerant_2001}

According to the Eqn. \ref{eqn:12}, the values of $\frac{ac}{b^2}$ can be estimated as 310 and 72 for YFe$_2$Zn$_{20}$ and LuFe$_2$Zn$_{20}$ respectively ($M/H \sim \chi (T)$ at 10~kOe), which are much larger than the region of the existence of IEM, indicating that IEM may not exist.
Indeed, recent measurements on a part of the LuFe$_2$Zn$_{20}$ sample used for the magnetization data in Fig. \ref{FigYLu1M} in a pulse magnetic field up to 55~T at 0.3~K, show no evidence of metamagnetic transition.
In nearly FM materials, no evidence of IEM appears for TiBe$_2$,\cite{yamada_high-field_1998}, which also manifests a relative low value of $T_{max}$.
From these points of view, Y(Lu)Fe$_2$Zn$_{20}$ and TiBe$_2$ may represent the examples of NFFLs different from YCo$_2$ and LuCo$_2$.

This lack of an IEM sheds further light on the magnetic properties of the local moment bearing, RFe$_2$Zn$_{20}$ (R = Gd - Tm) compounds.
As shown before, all the members manifest 2nd order paramagnetic to ferromagnetic phase transitions.
This behavior is different from that seen in the RCo$_2$ (R = Gd - Tm) system: the magnetic phase transitions of R = Dy - Tm members for RCo$_2$ are 1st order whereas R = Gd and Tb members have 2nd order transitions\cite{duc_formation_1999}.
This difference is not difficult to explain in Landau theory: unlike Y(Lu)Co$_2$, the host of Y(Lu)Fe$_2$Zn$_{20}$ lack of ability to show IEM and therefore can not be induced to show metamagnetic transition by any molecular field associated with the $4f$ local moments.

\section{Summery}

RFe$_2$Zn$_{20}$ and RCo$_2$Zn$_{20}$ (R = Gd - Lu, Y) demonstrate diverse magnetic properties.
The conspicuous differences between these two related series are mainly associated with the conduction electron polarizability of the host (non-magnetic) compounds.
YFe$_2$Zn$_{20}$ and LuFe$_2$Zn$_{20}$ manifest similar,nearly ferromagnetic properties.
When the $4f$ local moments are embedded in this highly polarizable medium, the RFe$_2$Zn$_{20}$ (R = Gd - Tm) series shows highly enhanced FM ordering temperatures.
In contrast, YCo$_2$Zn$_{20}$ and LuCo$_2$Zn$_{20}$ manifest normal, Pauli paramagnetic behaviors.
In a related manner, GdCo$_2$Zn$_{20}$ and TbCo$_2$Zn$_{20}$ show low temperature AFM ordering, and the magnetic properties for RCo$_2$Zn$_{20}$ (R = Dy - Tm) are more strongly influenced by the CEF splitting of the R ions (the dominant energy scale).
The CEF coefficients determined for the Co series are consistent with the observed anisotropies of the Fe series, indicating that the CEF splitting of the R-ions is similar for the all Zn Frank-Kasper polyhedra that fully encompass the R-site.
On the other hand, YbFe$_2$Zn$_{20}$ and YbCo$_2$Zn$_{20}$ manifest different heavy Fermion behaviors. 

\begin{acknowledgments}
The authors thank J. Frederich for growing some of the compounds, L. Tan for Laue X-ray measurements, E. D. Mun, Y. Janssen and R. Prozorov for helpful discussions.
Ames Laboratory is operated for the U.S. Department of Energy by Iowa State University under Contract No. DE-AC02-07CH11358.
This work was supported by the Director for Energy Research, Office of Basic Energy Sciences. 

\end{acknowledgments}

\appendix
\section{}
\begin{figure}
  \begin{center}
  \includegraphics[clip, width=0.45\textwidth]{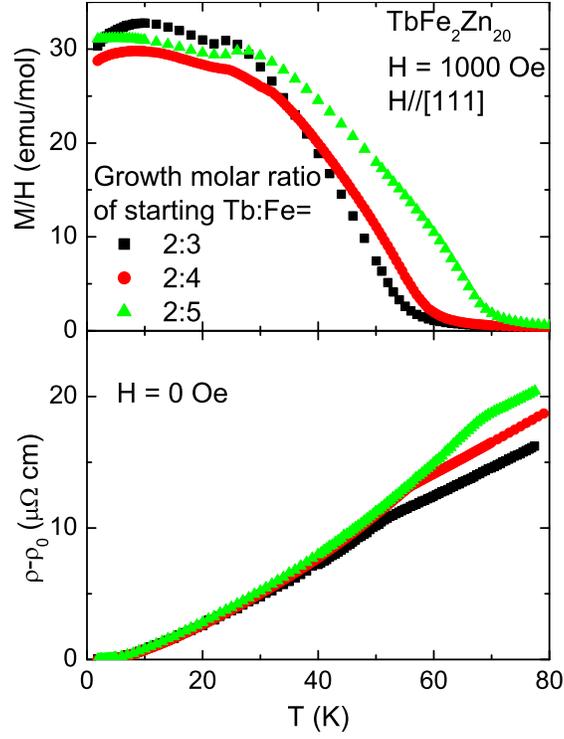}\\
  \caption{(Color online) (a): Temperature dependent $M/H$ for TbFe$_2$Zn$_{20}$ ($H = 1000$~Oe) from different initial growth molar ratio of starting elements; (b) temperature dependent $\rho$ in zero applied field.}
  \label{FigappenTb}
  \end{center}
\end{figure}

\begin{figure}
  \begin{center}
  \includegraphics[clip, width=0.45\textwidth]{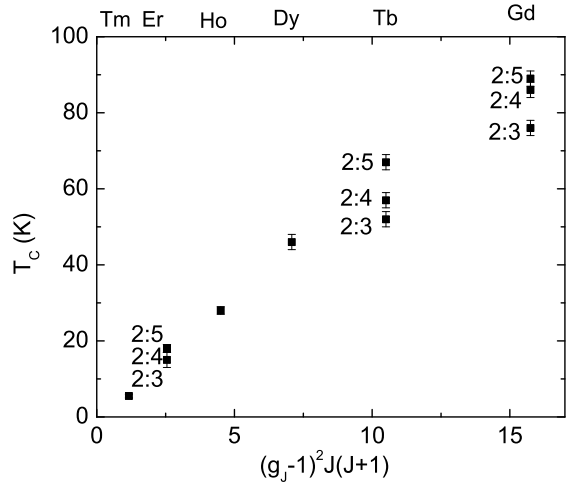}\\
  \caption{$T_{\mathrm{C}}$ values for different growth of RFe$_2$Zn$_{20}$ (R = Gd - Tm) samples with respect to the de Gennes factor.}
  \label{Figappen2}
  \end{center}
\end{figure}

Figure \ref{FigappenTb} shows the magnetization (at H = 1000 Oe) and zero applied field resistivity for three batches of TbFe$_2$Zn$_{20}$, which were synthesized from different initial molar ratio of starting elements, Tb:Fe:Zn = 2:3:95, 2:4:94 and 2:5:93. 
The ferromagnetic ordering temperatures, determined as $52\pm 2$~K, $56\pm 1$~K and $67\pm 2$~K by resistivity measurements for the three samples, increase as the initial concentration of Fe in the solution increases.
We also found similar features for R = Gd and Er, but the variation of their $T_{\mathrm{C}}$ values are less than in the Tb case (Fig. \ref{Figappen2}).
Comparative, single crystal x-ray diffraction measurements performed on the samples, albeit inconclusive, indicated that the crystallographic difference were perhaps due to subtle variations of occupancy of the Fe site.\cite{Xray}
The main difficulty associated with the x-ray diffraction measurement was completely resolving the mixed site occupancies for Zn and Fe which have similar atomic number values.
Recently, carefully prepared, pieces of TbFe$_2$Zn$_{20}$ samples with same geometric dimension, from the starting elements, Tb:Fe:Zn = 2:3:95 and 2:5:93, were measured using single-crystal neutron diffraction.\cite{neutron}
This measurement showed that the Fe site has $\sim 1${\%} deficiency for the 2:3:95 sample.  
All these crystallographic measurements indicate the sensitivity of the magnetic properties to the small disorder for RFe$_2$Zn$_{20}$ in this family which has local moments submerged in a highly polarizable conduction electron matrix.

\section{}
As shown in Fig. \ref{Fig1poly}, the distance between the rare earth ion and the Zn NNs, as well as NNNs, is close to 3~{\AA}; whereas the distance with the next next nearest neighbors (NNNN, 6 Zn in $48f$ site) is larger than 5~{\AA}.
Due to this isolated, cage-like coordinate of rare earth ions, the effect of ions other than this CN-16 Frank-Kasper polyhedron can be neglected in the calculation of the CEF coefficients, based on the point charge model.

Appreciably, the ligands of the rare earth ions in the C-15 Laves compounds (RNi$_2$) form a same polyhedron, whose CEF coefficients have been calculated by B. Bleaney \cite{bleaney_c15}, based on the point charge model. 
Therefore, one can directly cite the results:

\begin{equation}
B^0_4 = -\frac{3}{2}\left( \frac{91e^2Z_1}{726R^5_1}- \frac{7e^2Z_2}{54R^5_2}\right) \left\langle r^4\right\rangle \left\langle J \|\beta \| J\right\rangle
\label{eqn:A1}
\end{equation}

\begin{equation}
B^0_6 = \frac{9}{16}\left(- \frac{8e^2Z_1}{363R^7_1}- \frac{8e^2Z_2}{81R^7_2}\right) \left\langle r^6\right\rangle \left\langle J \|\gamma \| J\right\rangle,
\label{eqn:A2}
\end{equation}
where $Z_1e$ and $Z_2e$ are the charge of the NN and NNN ions ($Z_1=Z_2=2$ for Zn$^{2+}$), $R_1$ and $R_2$ is the distance between the R ion and the two sets of ions, $\left\langle r^4\right\rangle$ and $\left\langle r^6\right\rangle$ are the mean fourth and sixth powers of the electronic radius for the $4f$-electrons, and $\beta $ and $\gamma $ are the Steven multiplicative factors.
Extracting the values of $\left\langle r^4\right\rangle$ and $\left\langle r^6\right\rangle$ from ref.\cite{freeman_ion}, $\beta $ and $\gamma $ values from ref.\cite{lea_cef}, and $R_1$ and $R_2$ values from the results of single crystal X-ray diffraction\cite{Xray}, one can calculate the $B^0_4$ and $B^0_6$ values (shown in Table \ref{table3}). 


\end{document}